\documentclass[12pt]{amsart}

\usepackage{upgreek, amssymb, amsmath}
\usepackage[margin=3.5 cm]{geometry}
\usepackage{graphicx}
\usepackage{color}
\usepackage{subfigure}
\newtheorem{theorem}{Theorem}[section]

\newtheorem{coro}{Corollary}[section]

\newtheorem{conj}{Conjecture}[section]

\theoremstyle{definition}

\theoremstyle{remark}

\numberwithin{equation}{section}

\newcommand{\RR}{{\mathbb R}}

\newcommand{\ZZ}{{\mathbb Z}}
\newcommand{\QQ}{{\mathbb Q}}

\def\bm{\begin{pmatrix}}
\def\em {\end{pmatrix}}

\newcommand{\abs}[1]{\lvert#1\rvert}

\DeclareMathOperator{\im}{Im}
\newcommand{\ud}{\mathrm{d}}

\begin{document}

\title[]{Dynamics and spectral theory of quasi-periodic Schr\"odinger-type operators}

 %  Information for first author
\author{C. A. Marx}
  % Address of record for the research reported here
\address{Department of Mathematics, Oberlin College, Oberlin OH, 44074}
\email{cmarx@oberlin.edu}
 %  \thanks will become a 1st page footnote.
\author{S. Jitomirskaya}
\address{Department of Mathematics, University of California, Irvine CA, 92717}
\email{szhitomi@uci.edu}
\thanks{S.J. is a 2014-15 Simons Fellow. This research was partially supported by NSF DMS-1101578 and DMS-1401204.}

%    General info
%\subjclass[2000]{Primary 54C40, 14E20; Secondary 46E25, 20C20}

%\date{January 1, 2001 and, in revised form, June 22, 2001.}


%\keywords{Differential geometry, algebraic geometry}

\maketitle

\section{Introduction} \label{sec_intro}

Quasi-periodic Schr\"odinger-type operators naturally arise in solid state physics, describing the influence of an external magnetic field on the electrons of a crystal. In the late 1970s, numerical studies for the most prominent model, the almost Mathieu operator (AMO), produced the first example of a fractal in physics known as ``Hofstadter's butterfly'' \cite{Hofstadter_1976}, marking the starting point for the ongoing strong interest in such operators in both mathematics (several of B. Simon's problems \cite{Simon_15problems, Simon_2000}) and physics (e.g. Graphene, quantum Hall effect).

Whereas research in the first three decades was focused mainly on unraveling the unusual properties of the AMO and operators with similar structure of potential, in recent years a combination of techniques from dynamical systems with those from spectral theory has allowed for a more ``global,'' model-independent point of view. Intriguing phenomena first encountered for the AMO, notably the appearance of criticality corresponding to purely singular continuous (sc) spectrum for a measure theoretically typical realization of the phase, could be tested for prevalence in general models.

The intention of this article is to survey the theory of quasi-periodic Schr\"odinger-type operators attaining this ``global'' view-point with an emphasis on dynamical aspects of the spectral theory of such operators. For a more ``traditional'' review centered about the AMO we refer to e.g. \cite{Jitomirskaya_review_2007}. 

\section{Set up} \label{sec_models}

Fix $d, \nu \in \mathbb{N}$. A quasi-periodic $\nu$-frequency, matrix valued Schr\"odinger-type operator (on the line) is a family of self-adjoint, bounded operators on $\mathit{l}^2(\mathbb{Z}, \mathbb{C}^d)$ indexed by points $\theta \in \mathbb{T}^\nu:=\mathbb{R}^\nu / \mathbb{Z}^\nu$ of the form
\begin{eqnarray} \label{eq_operator_def}
[H_{\alpha,\theta} \psi]_n := C(T_\alpha^{n-1} \theta)^* \psi_{n-1} + C(T_\alpha^n \theta) \psi_{n+1} + V(T_\alpha^n \theta) \psi_n ~\mbox{.}
\end{eqnarray}
Here, for every realization of $\theta$, $H_\theta$ is generated by evaluating two matrix-valued sampling functions $C,V \in \mathcal{C}(\mathbb{T}^\nu, M_d(\mathbb{C}))$, $V = V^*$, along the rotational trajectory induced by $(T_\alpha \theta)_j = \theta_j + \alpha_j$, $1 \leq j\leq \nu$, where $\alpha \in \mathbb{T}^\nu$ is fixed with incommensurate components. $\theta$ is commonly called the {\it{phase}}, whereas $\alpha$ is referred to as the {\it{frequency}}. 

We emphasize that $C(\theta)$ in (\ref{eq_operator_def}) is not assumed to be invertible, in fact we call (\ref{eq_operator_def}) {\it{singular}} if $\det C(\theta) = 0$ for some $\theta \in \mathbb{T}^\nu$, and {\it{non-singular}} otherwise. For reasons of ergodicity we do however require that 
\begin{equation} \label{eq_integrabilitycond}
\log\vert \det C(.) \vert \in L^1(\mathbb{T}^\nu) ~\mbox{. } %\log \Vert C(.) \Vert \in L^1(\mathbb{T}^\nu) ~\mbox{.}
\end{equation}
Here, $\mathbb{T}^\nu$ is understood to be equipped with its Haar probability measure subsequently denoted by $\ud^\nu \theta$ (or $\vert . \vert$ as a set function).%; $\Vert . \Vert$ in (\ref{eq_integrabilitycond}) denotes any matrix norm on $M_d(\mathbb{C})$. 

Presently, most results are known for $d=\nu=1$, which will also be the main focus of this article. 
%We will signify the case $d=1$ choosing the lower case letters $c,v$ for the then scalar-valued sampling functions. 
The most widely studied special case arises when $d=1$ and $C \equiv 1$, 
\begin{equation} \label{eq_schrodingerop}
[H_{\alpha,\theta} \psi]_n = \psi_{n-1} + \psi_{n+1} + V(T_\alpha^n \theta) \psi_n ~\mbox{, } \psi \in \mathit{l}^2(\mathbb{Z},\mathbb{C}) ~\mbox{,}
\end{equation}
which is commonly known as quasi-periodic {\it{Schr\"odinger}} operator. Probably because the most prominent quasi-periodic operator, the AMO where $V(\theta)= 2 \lambda \cos(2 \pi \theta)$, $\lambda >0$, falls into this class, the majority of articles in the literature have focused on the special case (\ref{eq_schrodingerop}). 

From a physics point of view, however, the situation when $C \not \equiv 1$ or even has zeros is absolutely natural and cannot be excluded (cf. Sec. \ref{sec_models_physics}). To explicitly distinguish (\ref{eq_schrodingerop}) from the general case (\ref{eq_operator_def}), we shall from here on refer to the latter as (matrix-valued) {\it{Jacobi}} operator.

One of the contributions we hope to make with this review article, is to present a consistent picture for the general Jacobi case. Therefore, whenever the proof of a statement, originally in the literature for Schr\"odinger operators, has an obvious extension to the Jacobi case, we will give its Jacobi formulation. In some situations however, in particular when dealing with singular Jacobi operators, such extensions are not immediate and not in the literature, in which case we will state the theorems in their presently available Schr\"odinger form. In theses cases the authors hope to spark the interest in the reader to work on appropriate theorems applying to (singular) Jacobi operators.

%For $d=1$ important special case are {\em{Schr\"odinger}} operators where
%\begin{equation} \label{eq_schrodop}
%C \equiv \mathrm{I}_d ~\mbox{, } V = \begin{pmatrix} v(T_\alpha^{d-1} \theta) & 1 & \dots & 1 \\ 1 & \ddots & \ddots & \vdots \\ \vdots & \ddots & v(T_\alpha \theta) & 1 \\ 1 & \dots & 1 & v(\theta)   \end{pmatrix} ~\mbox{.}
%\end{equation} 
%The operators (\ref{eq_schrodop}) constitute the most widely studied model among the general form (\ref{eq_operator_def}). To explicitly distinguish (\ref{eq_schrodop}), we shall refer to the general form of (\ref{eq_operator_def}) as {\em{Jacobi}} operator. Though originally most results on quasi-periodic operators were proven for Schr\"odinger operators, we shall formulate them for the general Jacobi case whenever possible. Some extensions though, in particular when dealing with singular Jacobi operators are not immediate, in which case we will state the theorems in their original form. {\em{We will mainly focus on the case $d=\nu=1$ where presently most results are known}}. 

Finally, we mention that at times we will contrast results for Schr\"odinger operators on $\mathbb{Z}$ with their analogue on higher dimensional lattices, $\mathbb{Z}^N$ for $N>1$, where the first two terms in (\ref{eq_schrodingerop}) are correspondingly replaced by the $N$-dimensional discrete Laplacian. As the latter however do not lend themselves to the dynamical description central to this article, we will not describe the methods for their study here. An excellent reference for the Green's function methods suitable for quasi-periodic operators on higher dimensional lattices is e.g. \cite{Bourgain_book_2005}.

\subsection{Origin in physics} \label{sec_models_physics}
In physics, quasi-periodic Jacobi operators arise as effective Hamiltonians in a tight-binding description of a crystal subject to a weak external magnetic field. In particular, the case $d=\nu=1$ describes the effects for a two dimensional crystalline layer with $\alpha \in \mathbb{T}$ corresponding to the magnetic flux per unit cell exerted perpendicular to the lattice plane. Assuming a Bloch-wave like solution in one direction of the lattice plane, the conducting properties in the transversal direction are governed by (\ref{eq_operator_def}). In this context, $\theta$ is the quasi-momentum of the Bloch-wave, and the functions $C, V$, in applications trigonometric polynomials, reflect the lattice geometry and the allowed electron hopping between lattice sites. 

Even though the derivation of such models reaches back to the first half of the twentieth century \cite{Peierls_1933, Luttinger_1951, Harper_1955}, quasi-periodic Jacobi operators gained new relevance in relation with the integer quantum Hall effect as pointed out by Thouless et al \cite{ThoulessKohmotoNightingaleNijs_1982, AvronOsadchySeiler_2003}. We will comment on specific models, in particular the AMO (or Harper's model in physics literature), in Sec. \ref{sec_models}.

An important origin for Jacobi operators with $d>1$ is {\it{Aubry-Andr\'e duality}} \cite{AubryAndre_1980, AvronSimon_1983}. Considering (\ref{eq_operator_def}) for $d=\nu=1$, its dual is given by
\begin{eqnarray} \label{eq_dualop_jacobi}
[\widehat{H}_{\alpha, \theta} \psi]_n & := & \mathrm{e}^{-2 \pi i (\theta + \alpha n)} \left( \mathcal{F}^{-1} \overline{C} ~\ast ~\psi \right)_n \nonumber   \\
& & + ( \mathcal{F}^{-1} C ~\ast ~ \left( \mathrm{e}^{2 \pi i(\theta + \alpha m)} \psi_m \right)_{m \in \mathbb{Z}} )_n + (\mathcal{F}^{-1} V ~\ast ~ \psi)_n  ~\mbox{,}
\end{eqnarray}
where $\mathcal{F}: \mathit{l}^2(\mathbb{Z}) \to L^2(\mathbb{T})$ denotes the Fourier transform. If $C,V$ are trigonometric polynomials, the operator (\ref{eq_dualop_jacobi}) can be recast in the form (\ref{eq_operator_def}) where $d>1$ is the maximal degree of $C,V$ and the base dynamics is given by $(T_\alpha)^d$. In this case the $M_d(\mathbb{C})$-valued sampling functions take the form,
\begin{eqnarray}
V(\theta) &  =  & \begin{pmatrix} v(T_\alpha^{d-1} \theta) & \overline{c_1(T^{d - 2} \theta)} & \dots & \overline{c_{d-1}(\theta)} \\ c_1(T^{d - 2} \theta) & \ddots & \ddots & \vdots \\  \vdots & \ddots & v(T_\alpha \theta) & \overline{c_1(\theta)} \\ c_{d-1}(\theta) & \dots & c_1(\theta) & v(\theta) \end{pmatrix} ~\mbox{,} \\
C(\theta) & =  & \begin{pmatrix}  c_{d}(T_\alpha^{d-1} \theta) & c_{d-1}(T_\alpha^{d-1} \theta) & \dots & c_1(T_\alpha^{d-1} \theta) \\ 0  & \ddots & \ddots & \vdots \\ \vdots & \ddots & c_d(T_\alpha \theta) & c_{d-1}(T_\alpha \theta) \\ 0 & \dots & 0 & c_d(\theta)  \end{pmatrix} ~\mbox{,}
\end{eqnarray}
where
\begin{eqnarray}
c_j(\theta) = \overline{a_j} \mathrm{e}^{ - 2 \pi i \theta} + a_{-j} \mathrm{e}^{2 \pi i (\theta + j \alpha)} + \overline{b_j} ~\mbox{, } 1 \leq j\leq d ~\mbox{,} \\
v(\theta) = 2 \Re ( a_0 \mathrm{e}^{2 \pi i \theta} ) + b_0 ~\mbox{,}
\end{eqnarray}
for $V(\theta) = \sum_{\abs{j} \leq d} b_j \mathrm{e}^{2 \pi i \theta j}$ and $C(\theta) = \sum_{\abs{j} \leq d} a_j \mathrm{e}^{2 \pi i \theta j}$.

From a spectral theoretic point of view, the duality amounts to the generalized Fourier-transform,
\begin{equation} \label{eq_unit}
(\mathcal{U} \psi)(\eta, m):=\sum_{n \in \mathbb{Z}} \int_\mathbb{T} \mathrm{d} \theta \mathrm{e}^{2 \pi i m \theta} \mathrm{e}^{2 \pi i n (m \alpha+\eta)} \psi(\theta,n) ~\mbox{,}
\end{equation}
defined on the enlarged space $\int_\mathbb{T}^\oplus \mathit{l}^2(\mathbb{Z}) \mathrm{d} \theta$, mediating a unitary equivalence between the two decomposable operators with action on the fibers given by, respectively, $H_{\alpha, \theta}$ and $\widehat{H}_{\alpha,\theta}$ \cite{ChulDel}. Some spectral theoretic consequences of this unitary are given in Sec. \ref{sec_spectralprop_signatures}; an alternative dynamical approach to Aubry-Andr\'e duality will be discussed in Sec. \ref{sec_reducibility}.

\section{Basic spectral properties} \label{sec_spectralprop}

\subsection{Underlying structure} \label{sec_spectralprop_structure}

Letting $S$ denote the left shift on $\mathit{l}^2(\mathbb{Z},\mathbb{C}^d)$, the finite difference nature of (\ref{eq_operator_def}) implies the covariance relation
\begin{equation} \label{eq_covariance}
H_{\alpha, T_\alpha \theta} = S H_{\alpha,\theta} S^{-1} ~\mbox{,}
\end{equation}
whence all operators along the same rotational trajectory are unitarily equivalent. This underlying structure together with the properties of the base dynamics $T_\alpha$ leads to collective spectral behavior of the family of operators (\ref{eq_operator_def}) as indexed by the phase.

Ergodicity of $T_\alpha$ for incommensurate $\alpha$ makes (\ref{eq_operator_def}) an example of an {\it{ergodic operator}} thus the spectrum $\sigma(H_{\alpha,\theta})$ \cite{Pastur_1980} as well as its  Lebesgue components \cite{KunzSouillard_1980, KirschMartinelli_1982}, $\sigma_\gamma(H_{\alpha, \theta})$, $\gamma \in \{ac, sc, pp\}$, are for $\mu$-a.e. $\theta$ given by fixed compact sets, $\Sigma(\alpha)$ and $\Sigma_\gamma(\alpha)$, respectively. 

At this point we find it appropriate to mention two other heavily studied types of ergodic operators, the (discrete) {\it{Anderson model}} (or, more generally, (discrete) random Schr\"odinger operators) and {\it{Fibonacci Hamiltonians}} (or more generally, Schr\"odinger operators with dynamically defined potentials). Aside from ergodicity, however, most features as well as the techniques to study quasi-periodic models are very different from the aforementioned. In this article, we focus on the quasi-periodic case. In particular the dynamical systems approach to (\ref{eq_operator_def}) is either not particularly fruitful, as for random operators, or sufficiently different, as for Fibonacci Hamiltonians. For surveys of the Anderson model we refer to \cite{DisertoriKirschKleinKloppRivasseau_randomSchrodop_book_2008, Jitomirskaya_review_2007}, for Fibonacci Hamiltonians to \cite{Damanik_review2014}. 

Unique ergodicity of  $T_\alpha$ implies minimality, allowing for an approximation argument based on (\ref{eq_covariance}) and the continuity of $C,V$, which shows $\sigma(H_{\alpha,\theta})$ to be independent of $\theta$, for {\it{all}} $\theta \in \mathbb{T}^\nu$. Moreover, the discrete spectrum of $H_{\alpha, \theta}$ is empty for all $\theta \in \mathbb{T}^\nu$ \cite{Simon_ReviewAlmostPeriodic_1982}\footnote{The previous two statements on phase-insensitivity are obtained in \cite{Simon_ReviewAlmostPeriodic_1982} for all almost periodic operators.}. A more involved argument even proves constancy in $\theta$ of the absolutely continuous (ac) spectrum for a general minimal transformation $T$ \cite{LastSimon_1999}. We mention that for uniquely ergodic $T$ and Schr\"odinger operators with $d=1$, Kotani proved phase insensitivity of the ac spectrum using different means \cite{Kotani_1997}; while Kotani's result in \cite{Kotani_1997} is stated for continuous Schr\"odinger operators, it is easy to see that his proof extends to the discrete setup with $d=1$ discussed here. 

We emphasize that the everywhere independence of the ac spectrum of the phase does {\it{not}} carry over to the singular continuous (sc) or pure point (pp) component. A well known example is given by the supercritical AMO ($\lambda>1$) where the spectrum is purely sc for a dense $G_\delta$-set of phases \cite{JitomirskayaSimon_1994} and purely point outside this residual set \cite{Jitomirskaya_1999}, see also Theorem \ref{thm_jitosimon} and Sec. \ref{sec_models}. A famous conjecture, usually attributed to Simon, however proposes phase insensitivity of the {\it{singular spectrum}}, $\sigma_{sing}(H_{\alpha,\theta}):=\sigma_{sc}(H_{\alpha,\theta}) \cup \sigma_{pp}(H_{\alpha,\theta})$, i.e.
\begin{conj}[Simon] \label{conj_indepofsingularspec}
For all $d,v \in \mathbb{N}$ and all $\theta \in \mathbb{T}^\nu$, $\sigma_{sing}(H_{\alpha,\theta})$ is constant in $\theta$.
\end{conj}

Conjecture \ref{conj_indepofsingularspec} was first verified for the AMO for Diophantine $\alpha$ \cite{AvilaJitomirskaya_2010}; as a consequence of Avila proving the almost reducibility conjecture (ARC), it is now known more generally for $d=\nu=1$ and all non-singular Jacobi operators with {\it{analytic}} sampling functions $C,V$ and all irrational $\alpha$. We shall elaborate more on the ARC in Sec. \ref{sec_reducibility}. %\todo{\tiny{Confirm with Artur again whether we can formulate the ARC for non-singular Jacobi operators. On a previous account he mentioned that his proof is written for SL(2,$\mathbb{R}$) cocycles which, taking into account \cite{JitomirskayaMarx_preprint_2013}, would imply the ARC for the non-singular Jacobi case.}}

For singular Jacobi operators, $d=1$, and arbitrary $\nu$, Conjecture \ref{conj_indepofsingularspec} is in fact an immediate corollary to the following fact, which is interesting in its own right:
\begin{theorem} [\cite{Dombrowsky_1978}] \label{thm_dombrowsky}
For $d=1$, $\nu \in \mathbb{N}$ consider a {\it{singular}} Jacobi operator. Then for all $\theta \in \mathbb{T}^\nu$ one has $\sigma_{ac}(H_{\alpha,\theta}) = \emptyset$.
\end{theorem}
Theorem \ref{thm_dombrowsky} is another consequence of the minimality of $T_\alpha$ which, as $C$ is continuous, allows to a realize a given singular Jacobi operator for every $\theta$ as a trace-class perturbation of an infinite direct sum of {\it{finite-dimensional}} Jacobi blocks. 

From here on, {\it{if not specifically stated for a given result, the discussion will pertain to short-range ($d=1$), 1-frequency quasi-periodic Jacobi operators.}}

\subsection{Signatures of collective spectral properties} \label{sec_spectralprop_signatures}

Collectivity of spectral properties is expressed through the {\it{density of states}} (DOS) measure, defined by the spectral average 
\begin{equation} \label{eq_DOS}
n(\alpha; B) = \int_{\mathbb{T}^\nu} \mu_{\alpha,\theta}(B) \ud^\nu\theta ~\mbox{.}
\end{equation}
for a Borel set $B \subseteq \mathbb{R}$. Here, $\mu_{\alpha,\theta}$ is the spectral measure associated with $H_{\alpha,\theta}$ and the $\mathit{l}^2(\mathbb{Z})$ standard-basis vector $\delta_0$; in particular, $\mathrm{supp} (\ud n(\alpha)) = \Sigma(\alpha)$  \cite{AvronSimon_1983}. The definition (\ref{eq_DOS}) has an extension to $d>1$, see e.g. \cite{KotaniSimon_1988}.

The cumulative distribution $N(\alpha, E)$ of the DOS, commonly called {\it{integrated density of states}} (IDS), is known to be continuous \cite{JohnsonMoser_1982, AvronSimon_1983, DelyonSouillard_1984}, even $\log$-H\"older continuous \cite{CraigSimon_1983}. These statements hold true for arbitrary $d,\nu \in \mathbb{N}$; a proof for the modulus of continuity for $d>1$ can be found in \cite{HaroPuig_2013}. Recently, Klein and Bourgain \cite{BourgainKlein_2013} gave a {\it{deterministic}} proof of $\log$-H\"older continuity of the IDS which also extends to continuum Schr\"odinger operators in dimensions one, two, and three.  Note that above modulus of continuity cannot be improved in general: optimality was shown by Craig \cite{Craig_1983} and P\"oschel \cite{Poschel_1983}, see also \cite{KrugerZheng_2011} for a recent extension.
 %for the AMO square-root singularities are known to occur at the boundaries of gaps of the spectrum, see e.g. \cite{Puig_2006}. From a more general point of view, optimality of $\log$-H\"older continuity was shown for Schr\"odinger operators with limit periodic potentials in \cite{KrugerZheng_2011}.

We mention that an alternative perspective on the IDS, particularly suitable for dynamical considerations, comes through the {\it{rotation number}} \cite{JohnsonMoser_1982}. This dynamical route will arise naturally from the matrix cocycles associated with (\ref{eq_dualop_jacobi}) (see Sec. \ref{sec_dynamicalform_cocycles}) and will be discussed in Sec. \ref{sec_reducibility}.

Based on the spectral average (\ref{eq_DOS}) and the origin of quasi-periodic operators in physics (Sec. \ref{sec_models_physics}), it is natural to consider the direct integral operator $\mathcal{H}_\alpha:=\int_\mathbb{T}^\oplus H_{\alpha,\theta} \ud \theta$, mentioned before in context with Aubry-Andr\'e duality. It is easy to see that the spectrum of $\mathcal{H}_\alpha$ amounts to the {\it{union spectrum}} (see e.g. \cite{JitomirskayaMarx_GAFA_2012}), 
\begin{equation} \label{eq_splus}
S_+(\alpha):=\cup_\theta \sigma(\alpha,\theta) ~\mbox{.}
\end{equation} 
Related to the latter, one also introduces the {\it{intersection spectrum}}, 
\begin{equation}
S_-(\alpha):= \cap_\theta \sigma(\alpha, \theta) ~\mbox{.}
\end{equation}

From Sec. \ref{sec_spectralprop_structure}, irrationality of $\alpha$ implies $S_+(\alpha) = \Sigma(\alpha)$. The sets $S_\pm$ allow to approach quasi-periodic operators from periodic operators, approximating $\alpha$ by a sequence of rationals $(\frac{p_n}{q_n})$. Indeed $S_\pm$, first introduced by Avron, v. Mouche, and Simon \cite{AvronVMoucheSimon_1990}, played an important role in proving the Aubry-Andr\'e conjecture on the measure of the spectrum of the AMO (see Sec. \ref{sec_models}). 

As rational frequency approximation is indispensable for numerics, existence of the limits $S_\pm(\frac{p_n}{q_n})$ as $\frac{p_n}{q_n} \to \alpha$ is crucial. 
Questions of this form have a long history, particularly in context with the Aubry-Andr\'e conjecture; for a detailed survey including a list with references, see \cite{JitomirskayaMarx_GAFA_2012}. Results for general $C,V$ are much more sparse and depend on the regularity of the sampling functions and the topology underlying the limit.

Existence of limits in a set-wise sense modulo sets of zero Lebesgue measure is known in the analytic category, thereby allowing to recover the spectrum as well as its ac component through periodic approximation. Letting $\chi_B$ denote the characteristic function of a Borel set $B$ and $(\frac{p_n}{q_n})$ be the sequence of continued fraction approximants for a fixed irrational $\alpha \in \mathbb{T}$, one has:
\begin{theorem}[\cite{JitomirskayaMarx_GAFA_2012}\footnote{Ref. \cite{JitomirskayaMarx_GAFA_2012} proves Theorem \ref{thm_intersectionspectrum} for Schr\"odinger operators; the simple adaptations for the general Jacobi case are discussed in \cite{Marx_thesis}.}] \label{thm_intersectionspectrum}
Consider a non-singular Jacobi operator with $C,V$ {\it{analytic}}. Let $\alpha$ be a fixed irrational, then 
\begin{equation} \label{eq_thmintersectionspectrum}
\chi_{S_+(p_n/q_n)}(E) \to \chi_{\Sigma(\alpha)}(E) ~\mbox{, } \chi_{S_-(p_n/q_n)}(E) \to \chi_{\Sigma_{ac}(\alpha)}(E) ~\mbox{,}
\end{equation}
for Lebesgue a.e. $E$, in particular, $\vert S_+(p_n/q_n) \vert \to \vert \Sigma(\alpha) \vert$ and $\vert S_-(p_n/q_n) \vert \to \vert \Sigma_{ac}(\alpha) \vert$ as $\frac{p_n}{q_n} \to \alpha$.
\end{theorem}
Underlying Theorem \ref{thm_intersectionspectrum} are some basic properties of $S_+$, notably:
\begin{itemize}
\item[(a)] the invariance of $S_+(\alpha)$ under Aubry-Andr\'e duality for every $\alpha$ rational or irrational \cite{AvronSimon_1983, JitomirskayaMarx_GAFA_2012}. This property complements the invariance of the DOS measure under Aubry duality for irrational frequency, a simple consequence of the unitary (\ref{eq_unit}) \cite{GordonJitomirskayaGordonLastSimon_1997, JitomirskayaMarx_GAFA_2012}.
\item[(b)] continuity of $S_+(.)$ in the Hausdorff metric \cite{AvronSimon_1983}
\end{itemize}
Both (a) and (b) only require continuity of $C,V$; moreover, if $C,V \in \mathcal{C}^\gamma(\mathbb{T})$, $\gamma>0$ the modulus of continuity for (b) can be determined as $\frac{\gamma}{1+\gamma}$ \cite{JitomirskayaMavi_preprint_2012, AvronVMoucheSimon_1990}.

For Schr\"odinger operators with $V \in \mathcal{C}^\gamma(\mathbb{T})$, $\gamma > 1/2$, continuity of $S_+$ in the sense of (\ref{eq_thmintersectionspectrum}) was also proven upon restricting $S_+$ to the the set of positive Lyapunov exponent \cite{JitomirskayaMavi_preprint_2012}, see Theorem \ref{thm_Coro_positiveLE_Holder}. 

Addressing $S_-$, on the other hand, requires a detailed description of the set of zero Lyapunov exponent, which so far is only available for analytic sampling functions, cf. Sec \ref{sec_reducibility}. 

Based on a KAM scheme and relying on the spectral dichotomy described in Theorem \ref{thm_AFK}, Avila and Krikorian analyze the joint dependence of $S_\pm$ on the potential and the frequency. To this end, we say that for $\kappa, r >0$, $\alpha \in \mathbb{T}$ satisfies the {\it{Diophantine condition}} $DC(\kappa,r)$ if 
\begin{equation} \label{eq_dc}
\vert\vert\vert  q \alpha - p \vert\vert\vert > \kappa \vert q \vert^{-r} ~\mbox{, } (p,q) \in \mathbb{Z}^2 \setminus \{ (0,0) \} ~\mbox{.}
\end{equation}
Here, $\vert\vert\vert x \vert\vert\vert:=\inf_{n \in \mathbb{Z}} \vert x - n \vert$. It is well known that $\cup_{\kappa>0} DC(\kappa,r)$ has full measure provided $r>1$. 
\begin{theorem}[\cite{AvilaKrikorian_Annalspreprint_2006}]
Let $\kappa, r > 0$. The maps
\begin{equation}
(\alpha, V) \mapsto \vert \Sigma(\alpha) \vert ~\mbox{, } (\alpha, V) \mapsto \vert \Sigma_{ac}(\alpha) \vert ~\mbox{,}
\end{equation}
defined for quasi-periodic Schr\"odinger operators on $DC(\kappa,r) \times \mathcal{C}^\omega(\mathbb{T})$ are continuous.
\end{theorem}
We mention that the statement of the theorem appears in an unpublished version of the article of same title \cite{AvilaKrikorian_2006} which is available on Avila's webpage. Several other interesting conclusions about the continuity of various components of the spectrum, partly also valid for $V \in \mathcal{C}^\infty(\mathbb{T})$, can be found in Sec. 1.2 of \cite{AvilaKrikorian_Annalspreprint_2006}.

\subsection{Arithmetic conditions and singular continuous spectrum} \label{sec_spectralprop_distinguish}

%To conclude this section, we list some features distinguishing quasi-periodic operators from other ergodic operators, e.g. their random counterparts.

As mentioned earlier, quasi-periodic operators provide physical examples for the appearance of purely singular continuous spectrum. While the critical ($\lambda = 1$) AMO is the most well-known example for this intriguing phenomenon, more generally one can prove for quasi-periodic Schr\"odinger operators with {\it{continuous}} potentials\footnote{Theorem \ref{thm_aviladamanikbolshernitzan} is in fact proven for more general base dynamics, for details see \cite{AvilaDamanik_2005, BolshernitzanDamanik_2008}.}:
\begin{theorem}[\cite{AvilaDamanik_2005}; \cite{BolshernitzanDamanik_2008}] \label{thm_aviladamanikbolshernitzan}
Let $d=1$ and $\nu \in \mathbb{N}$ arbitrary. There is a dense $G_\delta$ set $\mathcal{SC} \subset \mathcal{C}(\mathbb{T}^\nu)$ such that for every $V \in \mathcal{SC}$ and a.e. $\theta \in \mathbb{T}^\nu$, the quasi-periodic Schr\"odinger operator $H_{\alpha,\theta}$ has purely sc spectrum.
\end{theorem}
The question of typicality of quasi-periodic Schr\"odinger operators with singular continuous spectrum has more recently been revisited for $V$ in the {\it{analytic}}  category (``Avila's global theory''). We describe these results in detail in Sec. \ref{sec_global} but emphasize at this point that Avila's analysis uses a different notion of typicality (``{\it{prevalence}},'') which is measure-theoretical instead of topological as in Theorem \ref{thm_aviladamanikbolshernitzan}.

%First,  we mention the possibility of purely sc spectrum for a {\it{measure-theoretically typical}} realization of $\theta$, which is in contrast to any known random operator. The prime example is the critical ($\lambda = 1$) AMO, however more generally one proves for quasi-periodic Schr\"odinger operators with continuous potential:
%\begin{theorem}[\cite{AvilaDamanik_2005}; \cite{BolshernitzanDamanik_2008}] \label{thm_aviladamanikbolshernitzan}
%Let $d=1$ and $\nu \in \mathbb{N}$ arbitrary. There is a dense $G_\delta$ set $\mathcal{SC} \subset \mathcal{C}(\mathbb{T}^\nu)$ such that for every $V \in \mathcal{SC}$ and a.e. $\theta \in \mathbb{T}^\nu$, the quasi-periodic Schr\"odinger operator $H_{\alpha,\theta}$ has purely sc spectrum.
%\end{theorem}

Appearance of eigenvalues in Theorem \ref{thm_aviladamanikbolshernitzan} is precluded by constructing Baire generic $V \in \mathcal{C}(\mathbb{T})$ which are ``close to periodic,'' so called {\it{Gordon-potentials}}. The latter name gives reference to a Lemma of A. Gordon \cite{Gordon_1976}\footnote{An English version of the proof can be found e.g. in \cite{CyconFroeseKirschSimon_book_1987}.} which provides a lower bound for the solutions to the Schr\"odinger equation for operators sufficiently close to periodic, whence precluding eigenvectors.

Gordon's Lemma also implies that the spectral properties of quasi-periodic operators depend sensitively on the arithmetic properties of the frequency: $\alpha \in \mathbb{T}$ is called {\it{$f$-Liouville}} if $\liminf_{n \to \infty} \frac{\vert\vert\vert \alpha - \frac{p_n}{q_n} \vert\vert\vert } { f( q_n)}<1$, for $f$ decaying  sufficiently fast  at infinity. Usually a particular $f$ is chosen when defining  Liouville numbers, but it is more convenient for us to use this more general definition here. For every $f>0$, $f$-Liouville numbers form a dense $G_\delta$. Following we use the convention that the {\it{point spectrum}} is the set of eigenvalues, whence the pure point spectrum $\sigma_{pp}$ is its closure.
\begin{theorem}[\cite{AvronSimon_1983}; \cite{Gordon_1976}] \label{thm_gordon}
Let $C, C^{-1},V \in \mathcal{C}(\mathbb{T}, M_d(\mathbb{C}))$, $d \in \mathbb{N}$.There exists a rate $f$ such that for every $f$-Liouville $\alpha$, $H_{\alpha,\theta}$ has empty point spectrum for all $\theta \in \mathbb{T}$. 
\end{theorem}
Notice that for singular operators one can never exclude eigenvalues for all $\theta$ due to possible appearance of finite Jacobi blocks for $\theta$ in the set
\begin{equation}
\mathfrak{S}(C):=\cup_{n \in \mathbb{Z}} T_\alpha^n( (\det C)^{-1}(\{ 0 \})) ~\mbox{.}
\end{equation}

We remark that the original formulation of Theorem \ref{thm_gordon} is for $d=1, C \equiv 1$ but the proof extends straightforwardly to the conditions of Theorem \ref{thm_gordon}. The theorem also holds for $\nu>1$, with an appropriate definition of $f$-Liouville.
Theorem \ref{thm_gordon} for singular Jacobi operators and a.e. $\theta$ cannot be obtained as an immediate adaptation of the classical proof. For $d=1$ it was proven in \cite{Koslover_2005}.

In addition to conditions on the frequency, arithmetic conditions on the phase are also in general necessary: Consider a quasi-periodic Schr\"odinger operator with potential  $V \in \mathcal{C}(\mathbb{T}^\nu)$, which is reflection symmetric about some point, expressed by $V(R\theta) = V(\theta)$, $\theta \in \mathbb{T}^\nu$. For $f\to\infty$, define the set of $f$-{\it{resonances}}, $\mathcal{R}(f):=\{ \theta : \liminf_{n \to \infty} \frac{\vert\vert\vert T_\alpha^{2 k} \theta - R \theta \vert\vert\vert }{f(k)}< 1\}$. As before, given $\alpha \in \mathbb{T}^\nu$ incommensurate, for each fixed $f>0$, $\mathcal{R}(f)$ forms a dense $G_\delta$. Assuming existence of an $\mathit{l}^2$-eigenvector $\psi$, $f$-resonances lead to an almost even potential, which would imply that $\psi$ is i.o. close to being even or odd; the later of course contradicts the decay of $\psi$, whence precludes existence of any $\mathit{l}^2$-eigenvectors. 
\begin{theorem} [\cite{JitomirskayaSimon_1994}] \label{thm_jitosimon}
%\footnote{In \cite{JitomirskayaSimon_1994} this Theorem is proven for the general almost periodic setup.}] \label{thm_jitosimon}
Let $V \in \mathcal{C}(\mathbb{T}^\nu)$ and $\alpha \in \mathbb{T}^\nu$ incommensurate. There exists $f=f(V)>0$ such that the Schr\"odinger operator $H_{\alpha, \theta}$ has empty point spectrum for all $\theta \in \mathcal{R}(f)$.
\end{theorem}
Theorem \ref{thm_jitosimon} extends without much change to the case of non-singular Jacobi operators with an appropriate symmetry of $C$. For singular Jacobi operators and $d=1$, the result was established in \cite{Koslover_2005}. 

\section{Dynamical Formulation} \label{sec_dynamicalform}

\subsection{Jacobi cocycles} \label{sec_dynamicalform_cocycles}

Given $\theta$, $E$ is called a {\it{generalized eigenvalue}} of $H_{\alpha,\theta}$ if the finite difference equation
\begin{equation} \label{eq_tiseq}
H_{\alpha,\theta} \psi = E \psi ~\mbox{, over } (\mathbb{C}^d)^\mathbb{Z} ~\mbox{,}
\end{equation}
admits a nontrivial solution with $\Vert \psi(n) \Vert \leq C(1+\vert n \vert)^\kappa$ for some $C,\kappa>0$, correspondingly called a {\it{generalized eigenvector}}. Denoting the set of generalized eigenvalues of $H_{\alpha,\theta}$ by $\mathcal{E}_g(H_{\alpha,\theta})$, it is well known that:
\begin{theorem}[\cite{Schnol_1957}; \cite{Berezanskii_1968}; \cite{Simon_1982}] \label{thm_schnol}
Let $d,\nu \in \mathbb{N}$ arbitrary. Given $\theta$,
\begin{equation} 
\sigma(H_{\alpha,\theta}) = \overline{\mathcal{E}_g(H_{\alpha,\theta})} ~\mbox{.}
\end{equation}
%Moreover, $\kappa$ can be restricted to at most $\frac{1}{2}+$. 
\end{theorem}
We mention that Theorem \ref{thm_schnol} holds more generally for a wide-class of self-adjoint operators, see \cite{KleinKoinesSeifert_2002} for more recent extensions and a list of references. A proof along the lines of \cite{Simon_1982} but specifically for discrete Schr\"odinger operators can be found e.g. in \cite{Kirsch_2007}.

Theorem \ref{thm_schnol} opens the door to a dynamical description of the spectral properties of (\ref{eq_operator_def}), studying the asymptotics of solutions to (\ref{eq_tiseq}). For $\nu, N \in \mathbb{N}$, the appropriate dynamical system is given by a {\it{matrix cocycle}}, a map defined on $\mathbb{T}^\nu \times \mathbb{C}^N$ by 
$(\alpha, D)(\theta, v):=(T_\alpha \theta, D(\theta) v)$, where $D: \mathbb{T}^\nu \to M_N(\mathbb{C})$ is measurable with $\log_+ \Vert D(.) \Vert \in L^1(\mathbb{T}^\nu)$ and $\alpha \in \mathbb{T}^\nu$ is fixed. In the context of Jacobi operators, $N=2 d$, see also (\ref{eq_transfer1}). We denote by $D_n(\alpha, \theta)$ the $n$th iterate of $(\alpha,D)$ on the fibers. A continuous cocycle $(\alpha, D)$ is called {\it{singular}} if $\det D(\theta) = 0$ for some $\theta \in \mathbb{T}^\nu$.

Cocycles are a suitable dynamical framework which capture the {\it{transfer matrix}} formalism for finite difference operators: Considering
\begin{eqnarray} \label{eq_transfer}
B^E(\theta):= \begin{pmatrix} C(\theta)^{-1} (E - V(\theta)) & - C(\theta)^{-1} C(T_\alpha^{-1} \theta)^* \\ I_d & 0_d \end{pmatrix} ~\mbox{,}
\end{eqnarray}
$(\alpha, B^E)$ iteratively generates solutions to (\ref{eq_tiseq}) by $B_n^E(\alpha,\theta) ( \begin{smallmatrix} \psi_0 \\ \psi_{-1} \end{smallmatrix}) = ( \begin{smallmatrix} \psi_n \\ \psi_{n-1} \end{smallmatrix})$. The choice of the transfer matrix (\ref{eq_transfer}) is not unique, e.g. following \cite{PasturFigotin_book, DamanikKillipSimon_2010}, 
\begin{equation} \label{eq_transfer1}
\widetilde{B}^E(\theta):= \begin{pmatrix} C(\theta)^{-1} (E - V(\theta)) & - C(\theta)^{-1} \\ C(\theta)^* & 0_d \end{pmatrix} ~\in Sp(d, \mathbb{C}) ~\mbox{,}
\end{equation}
yields a complex symplectic (this assumes $E \in \mathbb{R}$) and unimodular cocycle which relates to (\ref{eq_tiseq}) through $\widetilde{B}_n^E(\alpha,\theta) ( \begin{smallmatrix} \psi_0 \\ C(T_\alpha^{-1} \theta)^* \psi_{-1} \end{smallmatrix}) = ( \begin{smallmatrix} \psi_n \\ C(T_\alpha^{n-1} \theta)^*\psi_{n-1} \end{smallmatrix})$.

As $C(\theta)$ is in general not invertible for all $\theta$, singular and non-singular Jacobi operators can be treated on equal footing considering cocycles inheriting the regularity of $C,V$ (which however are singular in general): Based on (\ref{eq_transfer}) and Cramer's rule let \cite{Marx_preprint2013, KleinDuarte_2013_posLE}
\begin{equation} \label{eq_defAcoc}
A^E(\theta)= \begin{pmatrix}  \mathrm{adj}(C(\theta)) (E - V(\theta)) & - \mathrm{adj}(C(\theta)) C(T_\alpha^{-1} \theta)^* \\ \det C(\theta) I_d & 0_d \end{pmatrix} ~\mbox{,}
\end{equation}
where $\mathrm{adj}(M)$ is the adjugate of $M \in M_d(\mathbb{C})$ with the convention that $\mathrm{adj}(M) = 1$ if $d=1$. Similar to (\ref{eq_defAcoc}) one defines $\widetilde{A}^E(\theta)$ associated with (\ref{eq_transfer1}). Thus, whereas Schr\"odinger operators can be described in terms of continuous $SL(2,\mathbb{R})$ cocycles, for Jacobi operators consideration of possibly singular cocycles is unavoidable. $A^E$ and $\widetilde{A}^E$ are called {\it{Jacobi cocycles}} (and Schr\"odinger cocycle, if $C\equiv1$ and $d=1$).

The asymptotics of a matrix cocycle is described by the {\it{Theorem of Oseledets-Ruelle}} \cite{Oseledets_1968, Ruelle_1979}: An $M_N(\mathbb{C})$-cocycle $(\alpha, D)$ with $\alpha \in \mathbb{T}^\nu$ incommensurate, induces the invariant filtration $\{ 0 \} =V_{s+1} \subset V_s (\theta) \subset V_{s-1}(\theta) \subset \dots \subset V_1 = \mathbb{C}^N$ for some $1 \leq s \leq N$, such that $\frac{1}{n} \log \Vert D_n(\theta) v \Vert \to \lambda_j$ for $v \in V_{j}(\theta) \setminus V_{j+1}(\theta)$, a.e. $\theta$, and $1 \leq j \leq s$.  Note that if $(\alpha,D)$ is a singular cocycle, one has $\ker D(\theta) \subseteq V_1(\theta)$ in above filtration.

The numbers $\lambda_j$ repeated according to their multiplicity are called the {\it{Lyapunov exponents}}, $-\infty \leq L_N(\alpha,D) \leq \dots \leq L_1(\alpha,D)$ of the cocycle $(\alpha,D)$.
Since, $\sum_{k=1}^{N} L_k(\alpha,D) = \int_{\mathbb{T}^\nu} \log \vert \det D(\theta) \vert \ud^\nu \theta$, for $N=2$ the Lyapunov spectrum is completely described by the {\it{top Lyapunov exponent}}, 
\begin{equation} \label{eq_le}
L(\alpha,D) := L_1(\alpha) = \lim_{n \to +\infty} \frac{1}{n} \int_{\mathbb{T}^\nu} \log \Vert D_n(\alpha,\theta) \Vert \ud^\nu \theta ~\mbox{,}
\end{equation}
in which case we simply refer to it as {\it{the}} Lyapunov exponent  of the $M_2(\mathbb{C})$-cocycle $(\alpha,D)$. 

Existence of the various $a.e.$ limits relies on Kingman's sub-additive ergodic theorem. In view of rational frequency approximation, note that (\ref{eq_le}) is well-defined for {\it{all}} $\alpha \in \mathbb{T}^\nu$. Terminology-wise, as $(\alpha, B^E)$ directly relates to the generalized eigenvalue problem (\ref{eq_tiseq}), one usually refers to its LEs as the LEs of a given quasi-periodic Jacobi operator. 

Note that $(\alpha, A^E)$ and $(\alpha, \widetilde{A}^E)$ are measurably {\it{conjugate}}, in particular their Lyapunov spectra agree and
\begin{eqnarray}
L_j(\alpha, B^E) & = & L_j(\alpha, A^E) - \int_{\mathbb{T}^\nu} \log \vert \det C(\theta) \vert \ud^\nu\theta = L_j(\alpha, \widetilde{B}^E) ~\mbox{.}
\end{eqnarray}
Here, two cocycles $(\alpha, D)$ and $(\alpha, \widetilde{D})$ are called measurably conjugate if 
\begin{equation}
M(T_\alpha \theta)^{-1} D(\theta) M(\theta)= \widetilde{D}(\theta) ~\mbox{ a.e.}
\end{equation}
for some $M: \mathbb{T}^\nu \to M_d(\mathbb{C})$ measurable with $\log \Vert M(.) \Vert, \log \vert \det M(.) \Vert \in L^1(\mathbb{T}^\nu)$. This conjugacy together with the symplectic structure of $\widetilde{B}^E$ also implies that the LEs of a Jacobi operator counting multiplicity occur in positive-negative pairs. Observe that for non-singular Jacobi operators, all of the relevant cocycles are related by a conjugacy inheriting the regularity of $C,V$.

An important relation of the Lyapunov spectrum to the spectral properties of a quasi-periodic Jacobi operators is given by {\it{Thouless' formula}} \cite{Thouless_1972},
\begin{equation} \label{eq_thouless}
\gamma_+ = \int \log \vert E - E^\prime \vert \ud n(\alpha, E^\prime) - \frac{1}{d} \int_{\mathbb{T}^\nu} \log \vert \det C(\theta) \vert \ud^\nu\theta ~\mbox{,}
\end{equation}
Here, $\gamma_+$ is the sum of all non-negative LEs of $(\alpha, B^E)$ counting multiplicity. For $d=1$, (\ref{eq_thouless}) was established rigorously for ergodic Schr\"odinger operators in \cite{AvronSimon_1983, CraigSimon_1983}. An adaptation of the proof of \cite{CraigSimon_1983} for Jacobi operators with $d=1$ can be found e.g. in \cite{Teschl_book_2000}, respectively in \cite{KotaniSimon_1988} for $d>1$. A more dynamical approach for the special case of Jacobi operators with $d>1$ originating as dual operators to  quasi-periodic Schr\"odinger operators (cf (\ref{eq_dualop_jacobi})) was given in \cite{HaroPuig_2013}.

\subsection{Growth of transfer matrices}

Theorem  \ref{thm_schnol} precludes exponential growth of $\Vert B_n^E(\alpha, \theta)\Vert$ for all $\theta \in \mathbb{T}$ whenever $E \in \Sigma(\alpha)$. In fact, one may characterize the spectrum by the absence of such exponential growth: 

To formulate this precisely, given $N \in \mathbb{N}$, a {\it{continuous}} $M_N(\mathbb{C})$-cocycle $(\alpha, D)$ is said to induce a {\it{dominated splitting}} (write $(\alpha,D) \in \mathcal{DS}$) if it admits a {\it{continuous}}, invariant splitting of $\mathbb{C}^N = E_\theta^{(1)} \oplus  \dots \oplus E_\theta^{(s)}$ for some $2 \leq s \leq N$, such that for some $M \in \mathbb{N}$ and uniformly in $\theta$ one has $\Vert D_M(\alpha, \theta) w_j \Vert > \Vert D_M(\alpha, \theta) w_{j-1} \Vert$, for all $w_j \in E_\theta^{(j)}$ with $\Vert w_j \Vert=1$ and $2 \leq j \leq s$. $\mathcal{DS}$ specializes to {\it{uniform hyperbolicity}} ($\mathcal{UH}$) when considering unimodular continuous matrix cocycles. 
\begin{theorem}[\cite{Johnson_1986}; \cite{Marx_preprint2013}] \label{thm_johnson}
Let $d=1$ and $\nu \in \mathbb{N}$. $E \in \Sigma(\alpha)$ if and only if $(\alpha, A^E) \not \in \mathcal{DS}$. An analogous statement holds when $A^E$ is replaced by $\widetilde{A}^E$.
\end{theorem}
Johnson proved Theorem \ref{thm_johnson} for Schr\"odinger operators with $\mathcal{DS}$ replaced by $\mathcal{UH}$; a more recent, streamlined version of Johnson's proof can be found in \cite{Zhang_notes_2013}. An extension to Schr\"odinger operators with $d>1$ was established in \cite{HaroPuig_2013}. With a focus on singular Jacobi operators where correspondingly the notion $\mathcal{DS}$ becomes necessary, Theorem \ref{thm_johnson} was obtained in \cite{Marx_preprint2013}. In either case, the invariant subspaces are parametrized by the {\it{Weyl m-functions}}; for details we refer to e.g. Theorem 2.8 in \cite{HaroPuig_2013} for long-range Schr\"odinger, and Eq. (4.7)-(4.8) in \cite{Marx_preprint2013} for singular Jacobi operators (d=1).

An important strategy to prove presence of a $\mathcal{DS}$ is to establish existence of an {\it{invariant cone field}}, which also underlies the proof of Theorem \ref{thm_johnson} in \cite{Marx_preprint2013}. Detection in terms of invariant cone fields in particular implies that both $\mathcal{DS}$ and $\mathcal{UH}$ are open properties. A useful formulation of a cone field criterion suitable for $M_N(\mathbb{C})$-cocycles can be found e.g. in \cite{AvilaJitomirskayaSadel_2013}.

Complementing Theorem \ref{thm_johnson}, $\Sigma_{ac}(\alpha)$ may also be described in terms of the growth of transfer matrices; based on Theorem \ref{thm_dombrowsky} it suffices to consider non-singular Jacobi operators. 

Given a non-singular Jacobi operator, for $\theta \in \mathbb{T}^\nu$ consider the sets 
\begin{equation} \label{eq_lastsimonsets}
\mathcal{A}_\pm(\theta):=\{ E: \liminf_{n \to \pm \infty} \frac{1}{\vert n \vert} \sum_{j=1}^{\vert n \vert} \Vert B_j^E(\alpha,\theta) \Vert^2 < \infty \} ~\mbox{.}
\end{equation}
In \cite{LastSimon_1999}, Last and Simon\footnote{In \cite{LastSimon_1999}, Last and Simon consider Schr\"odinger operators, however the extension to non-singular Jacobi operators is immediate.} show that for all $\theta$, %$\mathcal{A}_+(\theta) \cup \mathcal{A}_-(\theta)$ forms an essential support of $\Sigma_{ac}(\alpha)$.
if $\mu$ is {\it{any}} spectral measure of $H_{\alpha,\theta}$, then $\mathcal{A}_+(\theta) \cup \mathcal{A}_-(\theta)$ forms an essential support of $\mu_{ac}$ and $\mu_{sing}$ is supported on the complement.

%Related to above characterization of $\Sigma_{ac}(\alpha)$, it is natural to conjecture that for a.e. $E$ w.r.t. the ac component of the spectral measure and for a.e. $\theta$ one even has $\displaystyle{\sup_{n \in \mathbb{Z}} \Vert B_n^E(\alpha, \theta) \Vert < \infty}$.  
Related to above characterization of $\Sigma_{ac}(\alpha)$, it is natural to conjecture that (at least for a.e. $\theta$), all generalized eigenfunctions of $H_{\alpha, \theta}$ are {\it{bounded}}, for a.e. $E$ w.r.t. to the absolutely continuous component of spectral measures.

For {\it{general}} uniquely ergodic base-dynamics,  this long-standing conjecture, sometimes called the {\it{Schr\"odinger conjecture}}, has recently been {\it{disproved}} by Avila \cite{Avila_SchrodingerConj_preprint}. In the quasi-periodic case with {\it{analytic}} sampling functions, the KAM scheme of \cite{AvilaFayadKrikorian_2011} on the other hand implies that the Schr\"odinger conjecture is true (cf. Sec. \ref{sec_reducibility}, Theorem \ref{thm_AFK}). Addressing the Schr\"odinger conjecture for quasi-periodic operators with lower regularities of the sampling functions still remains an open problem.

\subsection{Parameter complexification \& uniform domination} \label{sec_parametercomplexification}
\subsubsection{Energy complexification - Kotani theory} \label{Sec_paramtercomplexification_energy}
Theorem \ref{thm_johnson} characterizes the spectrum as a subset of the boundary of $\mathcal{DS}$. As mentioned earlier, $\mathcal{DS}$ and $\mathcal{UH}$ are stable properties accompanied by very regular dynamics, most notably continuity, even real analyticity of the LE \cite{Ruelle_1979_LEanalytic}. It is thus only natural to analyze $\Sigma(\alpha)$, approaching it from within $\mathcal{DS}$ by complexifying the energy.

From a dynamical point of view, {\it{Kotani theory}} \cite{Kotani_1982, Simon_1983} examines the limiting behavior of the invariant sections of $(\alpha, A^E)$ as $\Im E \to 0+$. We mention that while it is common to use the term Kotani theory, the discrete version was developed the discrete version for $d=1$ was developed by Simon \cite{Simon_1983}.

Being parametrized in terms of Weyl m-functions, existence of those limits rely on boundary values of Herglotz functions, consequently the derived statements are valid for {\it{Lebesgue a.e.}} $E$ and typically concern the set
\begin{equation}
\Sigma_0:=\{E: L(\alpha, B^E) = 0\} ~\mbox{,}
\end{equation}
shown to be essential support of $\Sigma_{ac}(\alpha)$ for any non-singular Jacobi operator. %Note that this characterization of the essential support of $\Sigma_{ac}(\alpha)$ implies that such limiting statements are a priori trivial for singular Jacobi operators since then $\vert \Sigma_0 \vert = 0$, as a consequence of Theorem \ref{thm_dombrowsky}. [CHECK THIS]
Kotani theory applies more generally to ergodic Schr\"odinger-type operators, both continuous and discrete, in particular the statements hold for all $\nu \in \mathbb{N}$. A detailed survey of Kotani theory and its spectral theoretic consequences for $d=1$ is given in \cite{Damanik_SimonFest_Kotani}.  Adaptations of the proofs to ergodic Jacobi operators and $d=1$ can be found e.g. in \cite{Teschl_book_2000}. \cite{KotaniSimon_1988} develops Kotani theory for $d>1$, proving that there is no odd multiplicity ac spectrum and the ac spectrum of multiplicity 2$j$ is supported on the set where exactly $2j$ of the LEs (counting multiplicity) equal zero. Extensions to unitary operators are discussed in \cite{Simon_OPUC_Part2} from the perspective of orthogonal polynomials on the unit circle.

For non-singular Jacobi operators and Lebesgue a.e. $E \in \Sigma_0$, the invariant sections possess limits in $L^2(\mathbb{T})$ as $\Im E \to 0+$. In fact, $(\alpha, A^E)$ is {\it{$L^2$-conjugate to a real (not necessarily constant) rotation}} which gives a precise formulation of the heuristics that solutions of (\ref{eq_tiseq}) for $E \in \Sigma_{ac}(\alpha)$ are determined by two Bloch waves propagating in opposite directions. A non-dynamical version of this result was proven for ergodic Schr\"odinger operators  by Deift and Simon \cite{DeiftSimon_1983}; the here presented dynamical formulation e.g. appears in \cite{AvilaKrikorian_2006}\footnote{In  \cite{AvilaKrikorian_2006}, the authors refer to H. Eliasson for pointing out to them the dynamical version of the original statement by Deift and Simon.}. A dynamical proof covering the Jacobi case is supplied in \cite{AvilaJitomirskayaMarx_preprint_2014}. 

It is an important question if such $L^2$-conjugacies lift to higher regularities depending on the sampling functions $C,V$. A $\mathcal{C}^r$ cocycle $(\alpha, A)$ with values in $SL(2,\mathbb{R})$, $r \in \mathbb{N} \cup \{0, \infty, \omega\}$, is called {\it{rotation-reducible}} if there exists $M \in \mathcal{C}^r(\mathbb{R}/2 \mathbb{Z},SL(2,\mathbb{R}))$ such that for all $\theta \in \mathbb{T}$
\begin{equation} \label{eq_rotred}
M(T_\alpha \theta)^{-1} A(\theta) M(\theta) \in SO(2,\mathbb{R}) ~\mbox{.}
\end{equation}
For later purposes, we emphasize that the definition of rotation-reducibility in (\ref{eq_rotred}) does {\it{not require a conjugacy to a constant rotation}}, the latter of which is usually called ``reducibility.'' Of course, for Diophantine $\alpha$ the two notions are equivalent by solution of a cohomological equation; as we will later be interested in statements for all irrational $\alpha$ (e.g. Theorem \ref{thm_AFK}), we clearly make this distinction.

Rotation-reducibility will be addressed in Sec. \ref{sec_reducibility} using KAM techniques. Since above mentioned $L^2$-conjugacy to rotations often serves as a crucial starting point in this context, the following sheds a light on the basic property underlying Kotani theory:
\begin{theorem}[\cite{AvilaKrikorian_monotonicty}] \label{thm_monotonicty}
Let $A_t \in \mathcal{C}^0(\mathbb{T},SL(2,\mathbb{R}))$ be a monotonic one-parameter family which is $\mathcal{C}^{2+\epsilon}$ in the parameter $t$. Then, for Lebesgue a.e. $t$ with $L(\alpha, A_t) = 0$, $(\alpha, A_t)$ is $L^2$-conjugate to (a not-necessarily constant) rotation.
\end{theorem}
We remark that the statement of Theorem \ref{thm_monotonicty} is proven more generally for cocycles where the base dynamics is given by a homeomorphism on a compact metric space.

Here, a parameter family $A_t$ is called {\it{monotonic}} if the argument of $t \mapsto A_t(x) \cdot v$ is positive for all $x \in \mathbb{T}$ and $v \in \mathbb{R}^2 \setminus \{0\}$. As argued in \cite{AvilaKrikorian_monotonicty}, monotonicity in the parameter $t$ implies $\mathcal{UH}$ once $t$ is complexified. In view of {\it{non-singular}} Jacobi operators, observe that even though $\widetilde{B}^E(x)$ is not monotonic itself, its second iterate $\widetilde{B}_2^E(x)$ is. We also mention that to apply Theorem \ref{thm_monotonicty} for Jacobi operators, by a standard unitary\footnote{The unitary replaces $C$ by $\vert C \vert$, see e.g. \cite{Teschl_book_2000}, Lemma 1.6. As shown in \cite{Marx_thesis} (see also \cite{AvilaJitomirskayaMarx_preprint_2014}), this unitary amounts dynamically to a measurable conjugacy, which, for a non-singular Jacobi operator with {\it{analytic}} $C$, inherits analyticity.} one may always assume $C$ to be real-valued.

For Jacobi operators with $d>1$, despite Kotani theory in its spectral theoretic formulation having been developed in \cite{KotaniSimon_1988}, a dynamical description of $\Sigma_{ac}(\alpha)$ is open. 

\subsubsection{Phase complexification} \label{sec_parametercomplexification_phasecomplexfication}
If the sampling functions $C,V$ are analytic, one may complexify the phase $\theta$, considering $A^E_y(\theta):=A^E(\theta + i y)$ for fixed energy $E \in \mathbb{R}$. Phase-complexification was originally proposed by Herman \cite{Herman_1983} and Sorets-Spencer \cite{SoretsSpencer_1991} to prove lower bounds on the LE for Schr\"odinger operators (see Sec. \ref{sec_supercritical_positiveLE}). 

In Avila's work on the global theory of one-frequency Schr\"odinger operators \cite{Avila_globalthy_published} (see also Sec. \ref{sec_global}), it became clear that phase-complexification can be employed to give an even finer description of the spectral properties of quasi-periodic Jacobi operators beyond what is known from Kotani theory. The approach is based on the behavior of the function $y \mapsto L(\alpha,A_y^E)$ in a neighborhood of $y=0$.  

The crucial property here is that for a every analytic cocycle $(\alpha, D)$, either $L(\alpha,D_y)= -\infty$ for all $y$ or $y \mapsto L(\alpha, D_y)$ is convex and piecewise linear with right-derivatives satisfying,
\begin{equation} \label{eq_acceleration}
\omega(\alpha,D_y):= \frac{1}{2 \pi} D_+ L(\alpha, D_y) \in \frac{1}{2} \mathbb{Z} ~\mbox{.}
\end{equation}
$\omega(\alpha,D_y)$ is called the {\it{acceleration of the cocycle}}, first introduced for $SL(2,\mathbb{C})$ cocycles, where it was shown to take only integer values (``{\it{quantization of the acceleration}}'') \cite{Avila_globalthy_published}. For singular cocycles, (\ref{eq_acceleration}) appears in \cite{JitomirskayaMarx_2013_erratum, Marx_thesis}, giving a criterion when the acceleration still remains integer valued, the latter of which being relevant for Jacobi operators. For higher dimensional cocycles, an appropriate generalization of quantization of the acceleration was obtained in \cite{AvilaJitomirskayaSadel_2013}.

In \cite{AvilaJitomirskayaSadel_2013}, the signature of $\mathcal{DS}$ is shown to be regularity of the cocycle, where $(\alpha,D)$ is called {\it{regular}} if $y \mapsto L(\alpha,D_y)$ is affine in a neighborhood of $y=0$:
\begin{theorem}[\cite{AvilaJitomirskayaSadel_2013}] \label{thm_dsAJS}
Let $\alpha$ be irrational. If $L_1(\alpha,D) > L_2(\alpha,D)$, $(\alpha,D) \in \mathcal{DS}$ if and only if it is regular. In particular, if $L(\alpha,D) > - \infty$, $(\alpha, D_y) \in \mathcal{DS}$ for $0<\vert  y \vert$ sufficiently small.
\end{theorem}
In \cite{AvilaJitomirskayaSadel_2013}, Theorem \ref{thm_dsAJS} is established for higher dimensional cocycles, in which case existence of gaps in the Lyapunov spectrum and an appropriate notion of regularity forms a necessary and sufficient criterion for domination.

Applied to Jacobi operators and taking into account Theorem \ref{thm_johnson}, the spectrum is partitioned into three mutually disjoint regimes according to the behavior of the {\it{complexfied LE}}, 
\begin{equation} \label{eq_defcomplexle}
L(E;y):= L(\alpha, A_y^E) - \int_\mathbb{T} \log\vert c(\theta) \vert \ud \mu(\theta)  \geq 0 ~\mbox{.}
\end{equation}
Observe that $L(E;y)$ is even in the parameter $y$ and $L(E;y=0)= L(\alpha, B^E)$.

For {\it{non-singular}} Jacobi operators, one hence distinguishes \cite{AvilaJitomirskayaMarx_preprint_2014, Avila_globalthy_published} between
\begin{enumerate}
\item {\it supercritical} energies, characterized by positive Lyapunov exponent (and thereby {\it{non-uniform domination}})
\item {\it subcritical} energies, characterized by the complexified Lyapunov exponent vanishing in  some neighborhood of $y=0$ (leading to almost-reducibility \cite{Avila_prep_ARC_1,Avila_prep_ARC_2}, see also Sec. \ref{sec_reducibility})
\item {\it critical} energies, characterized as being neither of the two above.
\end{enumerate}
We mention that the terminology was inspired by the AMO, where the three regimes are uniform over the spectrum, corresponding to the supercritical ($\lambda>1$), subcritical ($\lambda <1$), and critical $(\lambda=1$) AMO \cite{Avila_globalthy_published}.

Supercritical behavior is associated with Anderson localization (see Sec. \ref{sec_supercritical_localization}), whereas the subcritical regime supports only ac spectrum (see Sec. \ref{sec_reducibility}). Criticality gives rise to singular (sc+(possibly) pp) spectrum, see Sec. \ref{sec_reducibility}, Theorem \ref{thm_AFK}. For singular Jacobi operators, even though the notion of supercriticality still makes sense, distinguishing between subcritical and critical behavior does not provide additional insights due to absence of ac spectrum in the critical case (Theorem \ref{thm_dombrowsky}).

\section{Continuity of the Lyapunov exponent} \label{sec_continuity LE}

As mentioned earlier, presence of $\mathcal{DS}$ implies continuous, even real analytic dependence of $L(\alpha,D)$ on $(\alpha,D) \in \mathbb{R} \times \mathcal{C}(\mathbb{T},M_2(\mathbb{C}))$ \cite{Ruelle_1979_LEanalytic}. Since spectral theory however concerns the boundary of $\mathcal{DS}$, questions of continuity are delicate. 

While the LE is trivially upper-semicontinuous, lower-semicontinuity in general fails. First, continuous dependence on the frequency a priori excludes rational values, see e.g. \cite{Avila_globalthy_published} for a simple example of discontinuous behavior of the LE at rational $\alpha$. A natural example is also provided by the almost Mathieu operator \cite{Krasovksy_2000}. Second, questions of continuity in the matrix valued function depend strongly on the degree of regularity.

We also mention that for higher dimensional cocycles $N>1$, upper semicontinuity of the top LE generalizes to upper-semicontinuity of the maps $(\alpha,D) \mapsto \sum_{k=1}^r L_k(\alpha,D)$, for all $1\leq r \leq N$.

In the continuous category,  Furman \cite{Furmann_1997} showed that for fixed irrational $\alpha$, the top LE for a continuous $GL_N(\mathbb{C})$-cocycle $(\alpha,D)$ is discontinuous whenever the convergence of $\frac{1}{n} \log \Vert D_n(\alpha,\theta) \Vert \to L_1(\alpha,D)$ is non-uniform in $\theta$. The result is proven for base dynamics given by any uniquely ergodic, continuous homeomorphism $T$ on a compact metric space.  % As a crucial ingredient, Furman establishes uniformity of the {\em{upper}} limit of $\frac{1}{n} \log \Vert D_n(\alpha,\theta) \Vert$, proven more generally for averages of any sub-additive process, see also \cite{JitomirskayaMavi_preprint_2012} for a recent extension. Uniformity of the upper limits has important applications for Jacobi operators, in which context special cases of this result had been established earlier by Craig-Simon \cite{CraigSimon_1983} and Jitomirskaya \cite{Jitomirskaya_1999}. 

More generally, Bochi and Viana proved that for ergodic $T$, absence of $\mathcal{DS}$ and a non-trivial Lyapunov spectrum implies discontinuity of the Lyapunov exponents in the matrix-valued function \cite{BochiViana_2005}. This extended an earlier result obtained for $N=2$ in \cite{Bochi_2002}.

Discontinuous behavior of the LE in the continuous category is in strong contrast to the situation for analytic cocycles. Since analyticity is most natural in view of physical models for quasi-periodic Jacobi operators, questions of the continuity of the LE in the analytic category have a long history. The available results fall in one of two classes: 1. results on {\it{joint}} continuity in both frequency and matrix-valued function at any given irrational $\alpha$, and 2. a quantitative study of the modulus of continuity in the matrix valued function for {\it{fixed}} frequency\footnote{In principle the results of the second group do also admit varying the frequency, however only over a fixed Diophatine class. Rational frequency approximation of a given Diophatine $\alpha$ is thereby not possible.}. The second group of results depends on a quantitative description of the convergence in the subadditive ergodic theorem (``large deviation theorem''), whence imposing a Diophantine condition on $\alpha$ becomes necessary.

{\it{Among the first group, }} the most comprehensive result for $\nu=1$ was obtained in \cite{AvilaJitomirskayaSadel_2013} for analytic $M_N(\mathbb{C})$-cocycles:
\begin{theorem}[\cite{AvilaJitomirskayaSadel_2013}; \cite{JitomirskayaMarx_2012} for $N=2$] \label{thm_contiLE_AJS}
For $1 \leq k \leq N$, the Lyapunov exponents $\mathbb{R} \times \mathcal{C}^\omega(\mathbb{T}, M_N(\mathbb{C})) \ni (\alpha, D) \mapsto L_k(\alpha,D)$ are continuous at any $(\alpha^\prime, D^\prime)$ with $\alpha^\prime \in \mathbb{R}\setminus \mathbb{Q}$.
\end{theorem}
The proof of Theorem \ref{thm_contiLE_AJS} is based on phase-complexification and Theorem \ref{thm_dsAJS} (established also in \cite{AvilaJitomirskayaSadel_2013}), which yields $\mathcal{DS}$ accompanied by continuity of the Lyapunov exponents {\it{off}} the real axis \footnote{For completeness, we mention that \cite{AvilaJitomirskayaSadel_2013} proves Theorem \ref{thm_dsAJS} first in a weaker form, asserting $\mathcal{DS}$ off the real axis {\it for Lebesgue a.e.} $0< \vert y \vert$ sufficiently small. This weaker version is already enough to establish Theorem \ref{thm_contiLE_AJS}; it is in fact Theorem \ref{thm_contiLE_AJS} which then allows to drop the a.e. condition on $y$ which leads to the formulation of Theorem \ref{thm_dsAJS} given earlier.} Convexity of $y \mapsto \sum_{j=1}^{k} L_j(\alpha,D_y)$ implies that continuity persists when $y \to 0+$. Using a different strategy, related to the second group of continuity results mentioned above, Theorem \ref{thm_contiLE_AJS} was preceded by a result for analytic $SL(2, \mathbb{R})$-cocycles \cite{BourgainJitomirskaya_2002}. This result had later been refined by Bourgain, to prove joint continuity for analytic $GL_2(\mathbb{C})$-cocycles over rotations on $\mathbb{T}^\nu$ with $\nu \geq 2$ \cite{Bourgain_2005}. 

We stress that all the results in this first group in particular allow for rational approximations of $\alpha$, which is crucial for both practical (e.g. numerics) as well as theoretical considerations; for instance, Avila's global theory \cite{Avila_globalthy_published} (see also Sec. \ref{sec_global}) depends crucially on rational frequency approximation.

{\it{The second group of continuity results,}} quantify the modulus of continuity in the matrix valued function for positive LE and a fixed frequency, satisfying a {\it{strong Diophantine condition}} of the form $\vert \vert \vert n \alpha \vert \vert \vert \geq \frac{C}{n (\log n)^r}$ for some $r>1$. The strategy is based on two main ingredients, the Avalanche Principle and a suitable large-deviation theorem. 

The {\it{Avalanche Principle}} is a deterministic statement which allows to decompose large matrix products into smaller blocks, provided some largeness condition is met. Formulated by Goldstein and Schlag for $SL(2,\mathbb{C})$ matrices \cite{GoldsteinSchlag_2001}, a version suitable for $GL_N(\mathbb{C})$ can be found in \cite{DuarteKlein_2013_1}. We mention that similar ideas had been effectively used in \cite{young}, which later also inspired the work in \cite{Wang_You_2011} on discontinuity of the LE for smooth cocycles (see below). The {\it{large-deviation theorem}} quantifies the measure of the set $\{ \theta ~:~ \vert \frac{1}{n} \log \Vert D_n(\alpha,\theta) \Vert - L(\alpha,D) \vert \geq \delta \}$, outside of which the Avalanche Principle applies. 

The most comprehensive analysis so far has been carried out for $GL_N(\mathbb{C})$-cocycles by P. Duarte and S. Klein \cite{DuarteKlein_2013_1}, proving that if the Lyapunov spectrum has gaps, the gap pattern is locally stable and sums of Lyapunov exponents reflecting this gap pattern are H\"older continuous in the cocycle. The gap pattern in the Lyapunov-spectrum is the multi-dimensional analogue of the hypothesis of positivity of the LE for $d=1$ mentioned above. In particular, if all Lyapunov exponents are distinct, all Lyapunov exponents are H\"older continuous, a result which had been shown earlier in \cite{Schlag_2013}. The techniques also lend themselves to $\mathbb{T}^\nu$ and $\nu \geq 2$, resulting however in a modulus of continuity weaker than H\"older. Finally, we note that by Thouless' formula (\ref{eq_thouless}), H\"older continuity of $\gamma_+$ implies H\"older continuity of the IDS, thereby improving on its general $\log$-H\"older continuity in the regime where $\gamma_+>0$ (cf. Sec. \ref{sec_spectralprop_signatures}).

Most of the work in this second group of results has been done for Schr\"odinger like or more generally {\it{non}}-singular cocycles. Comparatively less is known for singular cocycles. For singular $M_2(\mathbb{C})$-cocycles, a large deviation theorem was obtained in \cite{JitomirskayaMarx_2011} under the constraint that $\det D(\theta)$ does not vanish identically\footnote{In this context, we mention that the continuity result in Theorem \ref{thm_contiLE_AJS} applies to all singular cocycles, even if the determinant vanishes identically.}. For singular Jacobi cocycles %\footnote{For singular Jacobi operators, the determinant of the Jacobi cocycle cannot vanish identically, otherwise $C \equiv 0$.} 
with analytic $C,V$, K. Tao later proved H\"older-continuity of the LE in the energy \cite{KTao_preprint_2011}. Recently, large deviation estimates covering not only the norm of the transfer matrices of a singular Jacobi operator but also its matrix elements have been proven in \cite{BinderVoda_2013}. Moreover, H\"older continuity of the IDS for singular, analytic quasi-periodic Jacobi operators on the set of positive LE and Diophantine frequency has recently been proven in \cite{taovoda}.

For higher-dimensional tori, a modulus of continuity on the matrix-valued function was obtained in \cite{KTao_2012}, however under certain restrictions on the zero-set of the determinant of $D$ in addition to requiring that $\det D \not \equiv 0$. 

Since the LE is continuous in the analytic, and discontinuous in the continuous category, it is a natural question whether the LE is continuous for intermediate regularities. Only recently this question has been answered negatively even if $D \in \mathcal{C}^\infty$! For singular cocycles, \cite{JitomirskayaMarx_2012} shows that if $\det D(\theta)$ is ``small and flat'' on a sufficiently large set (which is impossible for analytic functions), the LE is discontinuous at $D$ in $\mathcal{C}^k$ for $k \in \mathbb{N} \cup \{\infty\}$. 

A much more delicate argument by Y. Wang and J. You \cite{Wang_You_2011}, provides an example for discontinuity of the LE within the $SL(2, \mathbb{R})$, even within Schr\"odinger cocycles, in all $\mathcal{C}^k$ for $k \in \mathbb{N} \cup \{\infty\}$. As in \cite{JitomirskayaMarx_2012}, Wang and You's examples also rely on a lack of transversality on a ``large set'' which is possible in $\mathcal{C}^k$ but not in $\mathcal{C}^\omega$. 

Even though the examples in \cite{Wang_You_2011} were spectacular and surprising, one subtlety in these examples for discontinuity is that they require $\alpha$ to be a fixed irrational of {\it{bounded type}}, i.e. having a continued fraction expansion with bounded elements. The latter set includes the golden mean and is known to form a set of zero Lebesgue measure. This still leaves open the question whether continuous behavior of the LE at least for Schr\"odinger cocycles with $V \in \mathcal{C}^k$ is possible if $\alpha$  is not of bounded type. In this context we mention that for Schr\"odinger operators with Gevrey potentials, continuity of the LE in the energy is known if the potential satisfies a transversality condition and $\alpha$ is strongly Diophantine \cite{SKlein_2005}. Finally, the discontinuity construction in \cite{Wang_You_2011} has been modified to yield an even stronger statement: positivity of the Lyapunov exponent is not an open condition in $C^k$ \cite{Wang_You_2015}.

We also mention that while above examples for discontinuity are constructed in the Schr\"odinger cocycle, it is still an open question whether for a {\it{fixed}} potential of regularity lower than analytic, the LE is continuous or discontinuous in the {\it{energy}}.

\section{Rotation-reducibility} \label{sec_reducibility}
The connection between rotation-reducibility and absolutely continuous spectrum has been mentioned earlier in Sec \ref{Sec_paramtercomplexification_energy}. From (\ref{eq_lastsimonsets}), it is clear that showing rotation-reducibility (in at least the continuous category) for a positive measure set $\mathcal{S}$ of energies, implies that the spectrum is ac on $\mathcal{S}$. KAM theory provides a strategy for establishing rotation-reducibility, at least for cocycles of sufficient regularity. Most of the present section will be mainly concerned with Schr\"odinger operators with analytic $V$, even though some results appropriately generalize to non-singular Jacobi operators with smooth sampling functions. 

As common, we will use the terminology ``$D$ is {\it{reducible}},'' if $D$ is rotation-reducible to a {\it{constant}}; as mentioned earlier (see the comments after (\ref{eq_rotred})), it will however later be important to distinguish this special case from general rotation-reducibility defined in (\ref{eq_rotred}).

Following, let $(\alpha,D)$ be a matrix cocycle, $D \in \mathbb{C}^\omega(\mathbb{T}^\nu, SL(2,\mathbb{R}))$, which is homotopic to the identity. Obviously Schr\"odinger cocycles are homotopic to the identity. Nonsingular Jacobi operators also fall into this class, see \cite{AvilaJitomirskayaMarx_preprint_2014} for the necessary reductions. KAM schemes for cocycles non-homotopic to a constant have recently been developed by Avila and Krikorian in \cite{AvilaKrikorian_monotonicty}. 

In its simplest form, KAM theory allows to prove statements of the form, ``if $D$ is a small perturbation of a constant $A \in SO(2,\mathbb{R})$, then $D$ is reducible to $R_\rho$.'' Here, $\rho$ is the {\it{fibered rotation number}} of $D$, defined via a continuous lift $\tilde{F}: \mathbb{T}^\nu \times \mathbb{R} \to  \mathbb{T}^\nu \times \mathbb{R}$ of the map $(\theta, v) \mapsto (\theta + \alpha, \frac{D(\theta) v}{\Vert D(\theta) v \Vert})$ on $\mathbb{T}^\nu \times S^1$. Naturally, any such lift $\tilde F$ can be written as $\tilde{F}(\theta, x) = (\theta + \alpha, x + f(\theta,x))$, for some continuous $f$ satisfying $f(\theta, x+1) = f(\theta, x)$. The fibered rotation number $\rho(\alpha,D)$ is then defined by the limit,
\begin{equation}
\rho(\alpha,D):= \lim_{n \to \pm \infty} \frac{1}{n} \sum_{k=0}^{n-1} f(\tilde{F}^k(\theta,x) ~(\mathrm{mod} 1)) \in \mathbb{T} ~\mbox{,}
\end{equation}
which is independent of the lift and converges uniformly in $(\theta,x)$ to a constant with continuous dependence on the cocycle \cite{JohnsonMoser_1982, Herman_1983, DelyonSouillard_1983}. It is important to note that small divisor problems in the KAM strategy will in general impose arithmetic conditions on both $\alpha$ and $\rho$. 

In the context of Schr\"odinger cocycles, KAM naturally applies to the small potential regime, viewed as a perturbation of the pure Laplacian ($V \equiv 0$), whose Schr\"odinger cocycle is trivially a constant rotation for all $E \in \Sigma$ except at boundary points. For Schr\"odinger cocycles the rotation number is related to the IDS by
\begin{equation*}
N(\alpha,E) = 1 - 2 \rho(\alpha,E) ~\mbox{,}
\end{equation*}
in particular $\rho(\alpha,E)$ increases in $E$ with values in $[-1/2,0]$. Moreover, by the {\it gap-labeling theorem } \cite{JohnsonMoser_1982, DelyonSouillard_1983}, see also \cite{Bellisard_1993} for various generalizations, $N(\alpha,E)$ is constant on the connected components of $\Sigma^c$ with values in $\alpha \cdot \mathbb{Z}^\nu+ \mathbb{Z}$. For non-singular Jacobi operators a proof of the gap-labelling theorem can be found in \cite{DelyonSouillard_1983}. 

Usage of KAM theory for Schr\"odinger operators in the small potential regime goes back to Dinaburg and Sinai \cite{DinaburgSinai_1975}, later extended by Moser and P\"oschel \cite{MoserPoschel_1984}. The works of \cite{DinaburgSinai_1975, MoserPoschel_1984} show that for $\alpha$ Diophantine and $\rho(\alpha,E)$ either Diophantine or rational w.r.t. $\alpha$, there exists a positive measure set of rotation numbers such that if $V$ is small enough in $\mathcal{C}^\omega$, the system is reducible. It is important to note that the smallness criterion on $V$ depended on the rotation number, thereby preventing full measure statements.

It was Eliasson in \cite{Eliasson_1992} who first proved reducibility for a {\it{full}} measure set of rotation numbers:
\begin{theorem}[\cite{Eliasson_1992}] \label{thm_eliasson}
Let $\alpha \in \mathbb{T}^d$ Diophantine and $V: \mathbb{T}^d \to \mathbb{R}$ analytic on the strip of width $\delta$. There exist $\epsilon=\epsilon(\delta, \alpha)$ such that for $\Vert V \Vert_\delta < \epsilon$ and Lebesgue a.e. $E$, the corresponding Schr\"odinger cocycle is analytically reducible. Moreover, the spectrum of $H_{\alpha,\theta}$ is purely ac for a.e. $\theta$.
\end{theorem}
Here, $\Vert V \Vert_\delta:= \sup_{\vert \im (z) \vert \leq \delta} \vert V(z) \vert$. The full measure condition on $E$ is characterized by $\rho(\alpha, E)$ being Diophantine or rational w.r.t. $\alpha$, but the smallness criterion quantifying the ``{\it{Eliasson perturbative regime}}'' is independent of the rotation number! Eliasson also studies the remaining zero measure set of energies not covered by Theorem \ref{thm_eliasson}, for which he proves at most linear growth of the norms of the transfer matrices.  We mention that Theorem \ref{thm_eliasson} was proven in \cite{Eliasson_1992} for continuous Schr\"odinger operators, but the argument carries over to the discrete case, some details of which can be found in \cite{HadjAmor_2009} (see also \cite{AvilaJitomirskaya_2010}, Appendix A). 

Since the smallness condition on $V$ depends on $\alpha$, Theorem \ref{thm_eliasson} is an example of a {\it{perturbative result}}. A {\it{non-perturbative}} analogue of Theorem \ref{thm_eliasson} was established in \cite{BourgainJitomirskaya_2002_inventiones} for $\nu=1$ with $V$ of the form, $V(\theta) = \lambda f(\theta)$, $f \in \mathcal{C}^\omega(\mathbb{T})$; we denote the associated Schr\"odinger operator by $H_{\lambda; \alpha, \theta}$. The approach is based on the correspondence between localization of the dual operator to $H_{\lambda; \alpha, \theta}$, given by
\begin{equation} \label{eq_dualop_cos}
[\widehat{H}_{\lambda; \alpha,\theta} \psi]_n = \lambda \left( f \ast \psi \right)_n + 2 \cos(2 \pi (\theta + n \alpha)) \psi_n ~\mbox{,}
\end{equation}
and reducibility in $\mathcal{C}^\omega$: If for some $\theta$ and $E$, $\widehat{H}_{\lambda; \alpha,\theta}$ admits a generalized eigenfunction $u \in \mathit{l}^1$, then $(\alpha, A^E)$ is reducible in $\mathcal{C}$ to a constant rotation by $\theta$. Establishing exponential decay of $u$ correspondingly yields a conjugacy in $\mathcal{C}^\omega$. 

Extending the non-perturbative proof of Anderson localization for the AMO \cite{Jitomirskaya_1999} (see also Sec. \ref{sec_supercritical_localization} and \ref{sec_models}), Bourgain and Jitomirskaya developed a strategy to prove localization for the {\it{long-range operator}} in (\ref{eq_dualop_cos}),  resulting in:
\begin{theorem}[\cite{BourgainJitomirskaya_2002_inventiones}] \label{thm_bourgainJito_nonpert}
Let $H_{\lambda; \alpha,\theta}$ be a quasi-periodic Schr\"odinger operator with $\alpha$ Diophantine, $V(\theta) = \lambda f(\theta)$, $\lambda \in \mathbb{R}$, and $f: \mathbb{T} \to \mathbb{R}$ analytic. There exists $0 < \lambda_0 = \lambda_0(f)$ (independent of $\alpha$) such that if $\vert \lambda \vert < \lambda_0$, $H_{\lambda; \alpha, \theta}$ has purely ac spectrum for a.e. $\theta$.
\end{theorem}
We mention that, relying on Theorem \ref{thm_bourgainJito_nonpert}, Puig \cite{Puig_2006} removed the frequency dependence of the smallness condition in Theorem \ref{thm_eliasson} for $\nu=1$. It is important to note though that in the multifrequency case $\nu \geq 2$,  non-perturbative results of the kind of Theorem \ref{thm_bourgainJito_nonpert} are in general {\it{impossible}}, see e.g. the earlier review paper \cite{Jitomirskaya_review_2007}. 

As mentioned Theorem \ref{thm_bourgainJito_nonpert} explores the duality-based correspondence between localization and reducibility. This ``qualitative duality'' however faces certain a priori limitations: By Eliasson, reducibility is excluded outside a zero-measure set of energies, whereas localization fails outside a dense $G_\delta$-set of resonant phases (cf. Theorem \ref{thm_jitosimon}). In particular, these restrictions allow to only prove statements about ac spectrum valid for Lebesgue {\it{a.e.}} phase. 

These limitations were removed in \cite{AvilaJitomirskaya_2010} by developing a {\it{quantitative version}} of duality based on the dual concepts of {\it{almost reducibility}} and {\it{almost localization}}. As before, the results in \cite{AvilaJitomirskaya_2010} require $\alpha$ to be Diophantine. 

A cocycle is called almost reducible it the {\it{closure}} of its conjugacy class contains a constant. Put differently, almost reducible cocycles are analytically conjugate to a cocycle in Eliasson's perturbative regime. In particular, establishing almost reducibility entails all the dynamical and spectral conclusions valid in this regime. We note that Eliasson's perturbative regime is known to {\it not} be invariant under conjugacy. The idea of reducing non-perturbative (global) to perturbative (local) results originated from an earlier work by Avila and Krikorian \cite{AvilaKrikorian_2006} (cf. Theorem \ref{thm_AFK} below). 

It is shown in \cite{AvilaJitomirskaya_2010} that almost reducibility of $H_{\alpha,\theta}$ is implied by almost localization of its dual operator, (\ref{eq_dualop_cos}). Here, a generalized eigenfunction $\psi$ of $\widehat{H}_{\alpha,\theta}$ is called almost localized if it decays exponentially away from a sparse set of resonances. It should be noted that the cosine potential in (\ref{eq_dualop_cos}) implies that these resonances are energy independent, which is crucial for the approach in \cite{AvilaJitomirskaya_2010}. 

Using ideas from the non-perturbative analysis in \cite{BourgainJitomirskaya_2002_inventiones}, \cite{AvilaJitomirskaya_2010} proves that for $\lambda_0$ as in Theorem \ref{thm_bourgainJito_nonpert} and  $\vert \lambda \vert < \lambda_0$, the dual operator of $H_{\lambda; \alpha,\theta}$ presents almost localization for {\it{all}} $E \in \Sigma$ and {\it{all}} $\theta \in \mathbb{T}$. The latter thus yields:
\begin{theorem}[\cite{AvilaJitomirskaya_2010}] \label{thm_avilajitom_almostred}
Under the conditions of Theorem \ref{thm_bourgainJito_nonpert} and for the same value of $\lambda_0 = \lambda_0(v)$, one has for all $\vert \lambda \vert < \lambda_0$ that:
\begin{itemize}
\item[(i)]  $H_{\lambda; \alpha,\theta}$ has purely ac spectrum {\it{for all}} $\theta \in \mathbb{T}$
\item[(ii)] Conjecture \ref{conj_indepofsingularspec}  on the constancy of singular spectrum holds true
\item[(iii)] The IDS is $\frac{1}{2}$-H\"older continuous
\end{itemize}
Moreover, there exist a set $\mathcal{V} \subseteq \mathcal{C}^\omega(\mathbb{T})$ of infinite codimension such that for all $v \not \in \mathcal{V}$ and all but countably many $\vert \lambda \vert < \lambda_0$, all gaps of the spectrum are open.
\end{theorem} 
We remark that $\lambda_0$ determined based on Theorem \ref{thm_avilajitom_almostred} yields the optimal result $\lambda_0=1$ for the AMO ($v(\theta) = 2 \cos(2 \pi \theta)$). For the regime of couplings in Theorem \ref{thm_avilajitom_almostred}, it is the tight dynamical bounds in \cite{AvilaJitomirskaya_2010} establishing almost reducibility which imply $\frac{1}{2}$-H\"older continuity of the IDS (Lipschitz at reducible energies). We also mention that $\frac{1}{2}$-H\"older continuity of the IDS in Eliasson's perturbative regime had been proven independently % by Ben Hadj Amor
in \cite{HadjAmor_2009}. 

Supported by facts known for the AMO, this $\frac{1}{2}$-H\"older continuity is optimal in several ways: For the AMO, square-root singularities occur at the boundaries of the gaps of the spectrum, see e.g. \cite{Puig_2006}; moreover, for fixed $\vert \lambda \vert > 0$ and Baire-generic $\alpha$, the IDS of the AMO is not H\"older. Finally, for the critical AMO, even mild Diophantine conditions are not enough to guarantee H\"older continuity of the IDS \cite{Bourgain_book_2005}. 

From a global perspective, the spectrum may thus be partitioned into three regimes according to the dynamics of $(\alpha, A^E)$: 1. positive LE of the Jacobi operator, therefore non-uniform domination ($\mathcal{NDS}$), 2. almost reducibility ($\Sigma_{ar}$), and 3. neither of the above ($\Sigma_{crit}$). From \cite{AvilaJitomirskaya_2010}, if $\alpha$ is Diophantine, the spectrum is purely ac on $\Sigma_{ar}$ for all phases. That the same conclusion also holds for non-Diophantine $\alpha$ is more delicate, and was first proven in \cite{Avila_prep_ARC_1} for exponentially Liouville $\alpha$ and a.e. $\theta$, later extended to cover all irrational $\alpha$ and all $\theta \in \mathbb{T}$ in \cite{Avila_prep_ARC_2}. 

Since almost reducibility implies subexponential growth of $\Vert A_n^E(\theta) \Vert$ uniformly in $\theta$ across a band $\vert \im \theta \vert < \epsilon$, almost reducibility implies subcritical behavior in the sense of Sec. \ref{sec_parametercomplexification_phasecomplexfication}. The converse statement that subcriticality implies almost reducibility, is known as the {\it almost reducibility conjecture} (ARC), first formulated in \cite{AvilaJitomirskaya_2010}. 

The ARC was first verified for the AMO \cite{Avila_2008, AvilaJitomirskaya_2010}. In general, a proof of the ARC is announced by Avila in \cite{Avila_globalthy_published}, to appear in \cite{Avila_prep_ARC_2}. This extends the earlier result from \cite{Avila_prep_ARC_1}, where the conjecture has been proven for exponentially Liouvillean $\alpha$. In summary, above partitioning of the spectrum into $\mathcal{NDS}$,  $\Sigma_{ar}$, and $\Sigma_{crit}$ thus agrees with the partitioning into, respectively, supercritical, subcritical, and critical energies introduced in the end of Sec. \ref{sec_parametercomplexification_phasecomplexfication}, establishing a full correspondence between spectral/dynamical and analytical properties.

Whereas subcritical behavior is a signature of ac spectrum, the works of Avila, Krikorian, and Fayad \cite{AvilaKrikorian_2006, AvilaFayadKrikorian_2011} show that the spectrum is purely singular (sc+pp) on $\Sigma_{crit}$. The statement, which holds more generally for non-singular Jacobi operators \cite{AvilaJitomirskayaMarx_preprint_2014}, is based on the following spectral dichotomy, implying that $\vert \Sigma_{crit} \vert =0$:
\begin{theorem}[\cite{AvilaKrikorian_2006}, \cite{AvilaFayadKrikorian_2011}] \label{thm_AFK}
Consider a non-singular Jacobi operator with analytic sampling functions and $\alpha$ irrational. Then,  Lebesgue a.e. $E \in \Sigma(\alpha)$ is either rotation-reducible or belongs to $\mathcal{NDS}$.
\end{theorem}
Theorem \ref{thm_AFK} was first proven for $\alpha$ satisfying a recurrent Diophantine condition\footnote{$\alpha$ is called recurrent Diophantine if infinitely many iterates of it under the Gauss map satisfy a fixed Diophantine condition (\ref{eq_dc}).} \cite{AvilaKrikorian_2006}, and later extended in \cite{AvilaFayadKrikorian_2011} to all irrational $\alpha$. By Theorem \ref{thm_johnson}, Theorem \ref{thm_AFK} follows by showing reducibility (to constants for Diophantine $\alpha$) for Lebesgue a.e. $E$ with $L(\alpha, B^E)=0$. The approach in \cite{AvilaKrikorian_2006, AvilaFayadKrikorian_2011} reduces global reducibility results to local (i.e. perturbative) ones, which is possible using a clever renormalization scheme, associating  cocycles with $\mathbb{Z}^2$-actions. The crucial input here comes from Kotani theory, which ensures that for Lebesgue a.e. $E$ with $L(\alpha, B^E) = 0$ the cocycle is $L^2$-reducible (see Sec. \ref{sec_parametercomplexification}). A {\it{differentiable rigidity theorem}} also established in \cite{AvilaKrikorian_2006, AvilaFayadKrikorian_2011} then ensures that if an analytic $SL(2,\mathbb{R})$ cocycle is $L^2$-reducible, it is already so in $\mathcal{C}^\omega$. We mention that for $\alpha$ recurrent Diophantine, \cite{AvilaKrikorian_2006} obtains Theorem \ref{thm_AFK} also for $\mathcal{C}^\infty$ sampling functions. Finally, in the more general context of Theorem \ref{thm_monotonicty}, the differentiable rigidity theorem has recently been generalized to also hold for $SL(2,\mathbb{R})$-cocycles of class $\mathcal{C}^r$, $r \in \{ \infty, \omega \}$, which are non-homotopic to a constant (see Theorem 1.8 in \cite{AvilaKrikorian_monotonicty}).

We emphasize that while Theorem \ref{thm_AFK} implies that all spectral measures are purely singular on $\Sigma_{crit}$, the {\it{critical energy conjecture}} claims absence of point spectrum on $\Sigma_{crit}$, thereby:
\begin{conj}[Critical energy conjecture (CEC) \cite{AvilaJitomirskayaMarx_preprint_2014}] \label{conj_cec}
Let $\alpha$ be irrational. For a non-singular, quasi-periodic Jacobi operator with analytic sampling functions, the spectrum on $\Sigma_{crit}$ is purely sc for all $\theta \in \mathbb{T}$. For singular Jacobi operators, the same holds true for the set $\{E: L(E)=0\}$ and a.e. $\theta \in \mathbb{T}$.
\end{conj}
For singular Jacobi operators, exclusion of a zero measure set of phases in the CEC is indeed necessary, as zeros in $C$ may lead to finite Jacobi blocks \cite{AvilaJitomirskayaMarx_preprint_2014}. 

Establishing the CEC would yield the long sought-after {\it{direct}} criterion for detecting presence of singular continuous spectrum for quasi-periodic Jacobi operators, at least if the sampling functions are analytic. In  \cite{AvilaJitomirskayaMarx_preprint_2014} the CEC for a.e. $\theta \in \mathbb{T}$, has recently been proven for extended Harper's model, a quasi-periodic Jacobi operator generalizing the AMO (see Sec. \ref{sec_models}). %For the AMO, this has in fact been obtained earlier in an unpublished note by Avila \cite{Avila_preprint_2008_2}; in particular, it implies that the critical AMO has purely sc spectrum for a.e. $\theta \in \mathbb{T}$. 

Finally, we mention that the CEC can be considered a special case of a problem posed by Damanik in \cite{Damanik_SimonFest_Kotani}), asking to prove or disprove that for ergodic Schr\"odinger operators, the set of zero LE does not contain any eigenvalues.

\section{Super-critical regime and localization} \label{sec_supercritical}

\subsection{Positivity of the Lyapunov Exponent} \label{sec_supercritical_positiveLE}

In physics literature positivity of $L(\alpha, B^E)$ is often taken as an implicit definition of localization, and the Lyapunov exponent is often called the inverse localization length. If $L(\alpha, B^E) > 0$ for all $E\in \mathbb{R},$ there is no absolutely continuous component in the spectrum for all $\theta$ (see e.g. \cite{CyconFroeseKirschSimon_book_1987} for this result known as the Pastur-Ishii theorem). Positivity of the Lyapunov exponent, however, does not imply localization or exponential decay of eigenfunctions (in particular, not for Liouville frequency (Theorem \ref{thm_gordon}) nor for the resonant $\theta \in \mathbb{T}^\nu$ specified in Theorem \ref{thm_jitosimon}).

Non-perturbative methods to prove localization, at least in their original form, stem to a large extent from estimates involving the Lyapunov exponent and exploiting its positivity. The model considered in many results on positivity is a quasi-periodic Schr\"odinger operator with potential given by $V(\theta) = \lambda f(\theta)$ and coupling constant $\lambda \in \mathbb{R}$, in which context, the general theme is that $L(\alpha, B^E) > 0$ for sufficiently large $\lambda$. 

This subject has a rich history. A first proof was given by Herman \cite{Herman_1983} for a Schr\"odinger operator with $V$ given by a trigonometric polynomial, exploiting the subharmonicity of $\frac{1}{n} \log \Vert A_n^E(\alpha, \theta) \Vert$ in the complexified phase $\theta$. Herman's lower bound was in terms of the highest order coefficient of the trigonometric polynomial and therefore did not easily extend to the real analytic case. Subsequent proofs, however, were also based on subharmonicity.

Sorets-Spencer \cite{SoretsSpencer_1991} proved that for non-constant real analytic $f$ on $\mathbb{T}$ one has $L(\alpha, B^E) > \frac{1}{2} \log \vert \lambda \vert$ for $\vert \lambda \vert > \lambda_0(f)$ and all irrational $\alpha$. Another proof was given in \cite{BourgainGoldstein_2000}, where this was also extended to the multi-frequency case ($\nu>1$) with, however, the estimate on $\lambda_0$ also depending on the Diophantine condition on $\alpha$. Finally, Bourgain \cite{Bourgain_2005} proved continuity of the Lyapunov exponent in $\alpha$ at every incommensurate $\alpha$ (for $\nu>1$; for $\nu=1$ this was previously established in \cite{BourgainJitomirskaya_2002}, see Sec. \ref{sec_continuity LE}), which led to the following final statement:
\begin{theorem}[\cite{Bourgain_2005}] \label{thm_lyapbourg}
Consider a quasi-periodic Schr\"odinger operator $H_{\alpha, \theta}$ with $V(\theta) = \lambda f(\theta)$, where $f$ is a non-constant real analytic function on $\mathbb{T}^\nu$ and $\alpha \in \mathbb{T}^\nu$ is incommensurate. Then, there exists $\lambda_0 = \lambda_0(f)>0$ such that if $\vert \lambda \vert > \lambda_0$ one has $L(\alpha, B^E) > \frac{1}{2} \log \vert \lambda \vert$ for all $E \in \mathbb{R}$ and all incommensurate $\alpha$.
\end{theorem}

The main idea in the proof of Theorem \ref{thm_lyapbourg} is to use real analyticity of $f$ to achieve that  $(E- V(\theta))$ in the Schr\"odinger cocycle is bounded away from zero, at least outside a set of controllably small measure. Here, real analyticity enters through the {\it{Lojasciewicz' inequality}} which allows to estimate $\vert \{ \theta \in \mathbb{T}^\nu ~:~ \vert E - f(\theta) \vert < \delta \} \vert < \delta^c$ for $c=c(f)$ independent of $E$.

For $\nu =1$, the argument is even simplified based on the discreteness of the of zero set for non-constant complex analytic functions in one variable. Complexifying the phase implies existence of $y_0>0$ such that $(E - V(\theta + i y_0))$ is bounded away from zero uniformly in $(\theta, E)$, which in turn yields Theorem \ref{thm_lyapbourg} for the complexified LE at $y_0>0$. While Bourgain's proof of Theorem \ref{thm_lyapbourg} uses harmonic measure estimates to extract from this the positivity at $y =0$ (see also \cite{Bourgain_book_2005}, Ch. 3), Zhang \cite{Zhang_2012} and Duarte-Klein \cite{KleinDuarte_2013_posLE} realized that the argument can be considerably simplified using convexity of the complexified LE\footnote{More generally, a well-known result due to Hardy (see e.g. \cite{Duren_Hpfunctions_book} Theorem 1.6, therein) asserts that angular averages of functions subharmonic on an annulus are $\log$-convex in the radial variable.}. 

Using this approach, Duarte-Klein extended Theorem \ref{thm_lyapbourg} for $\nu=1$ to matrix-valued, singular Jacobi operators:
\begin{theorem}[\cite{KleinDuarte_2013_posLE}] \label{thm_duarteklein_positiveLE}
Consider a matrix-valued quasi-periodic Jacobi operator ($d \in \mathbb{N}$,  $\nu =1$ in (\ref{eq_operator_def})) with $V(\theta) = \lambda D(\theta)$, $\lambda \in \mathbb{R}$, $D \in \mathcal{C}^\omega(\mathbb{T}, M_d(\mathbb{C}))$ real symmetric, and $C \in \mathcal{C}^\omega(\mathbb{T}, M_d(\mathbb{C}))$. Assume moreover that 
\begin{equation} \label{eq_transversalitycondDuarteKlein}
\det C(\theta) \not \equiv 0 ~\mbox{, } D(\theta) ~\mbox{has no constant eigenvalues.}
\end{equation}
There exist $0 < \lambda_0 = \lambda_0(C, D)$ and $\gamma=\gamma(C,D)$ such that if $\vert \lambda \vert > \lambda_0$, one has positivity of the $d$ largest Lyapunov exponents of the Jacobi operator with
\begin{equation}
L_k(\alpha, B^E) \geq \log \vert \lambda \vert - \gamma ~\mbox{,}
\end{equation}
for all $E \in \mathbb{R}$ and $1 \leq  k \leq d$.
\end{theorem}
In \cite{KleinDuarte_2013_posLE}, the transversality condition (\ref{eq_transversalitycondDuarteKlein}) is shown to be measure theoretically generic (or {\it{prevalent}}). We mention that, preceding Theorem \ref{thm_duarteklein_positiveLE}, some partial results for quasi-periodic long-range Schr\"odinger operators ($C \equiv I_d$) with diagonal matrix $V$ had been obtained earlier in \cite{GoldsheidSorets_1992}.

Similar to the situation in Sec. \ref{sec_continuity LE} - \ref{sec_reducibility}, results on positivity for regularities of the sampling functions lower than analytic are much more sparse. In \cite{SKlein_2013_preprint} (\cite{SKlein_2005} for $\nu=1$), S. Klein extended Theorem \ref{thm_lyapbourg} to $f$ given by a Gevrey function on $\mathbb{T}^\nu$ and strongly Diophantine $\alpha \in \mathbb{T}^\nu$. Here, $f$ is assumed to satisfy a transversality condition, which requires a non-vanishing derivative of some order at each point. The transversality condition allows to prove a Lojasiewicz-type estimate for multivariable smooth functions (Theorem 5.1 in \cite{SKlein_2013_preprint}). In contrast to Theorem \ref{thm_lyapbourg}, $\lambda_0$ obtained in \cite{SKlein_2013_preprint} depends on both $f$ as well as the constants in the Diophantine condition on $\alpha$.

In the perturbative setting, i.e. with results holding for $\vert \lambda \vert>\lambda_0(f,\alpha)$ or for $\alpha$ in a set of measure going to zero as $\lambda \to \infty$, positive Lyapunov exponents
are known for $\nu=1$ and $f \in C^2$ with $f$ of $\cos$-type (in a certain sense) \cite{Sinai_1987, FrlichSpencerWittwer_1990}. Removing the $\cos$-type condition in the smooth category has been a subject of significant efforts and presents a serious challenge in non-uniformly-hyperbolic dynamics. 

Positivity of the Lyapunov exponents for energies outside a set of small (going to zero superpolynomially fast as $\lambda \to \infty$) measure holds for general ergodic operators by a result of Spencer \cite{sp} (recently improved to even a superexponential rate by Spencer and Shamis \cite{ss}, under an additional assumption that holds for higherdimensional skew-shifts but not shifts). There are several results with energy exclusion specific for quasiperiodic potentials \cite{Bjerklov_2005,ChanGoldsteinSchlag_prepr2006}, but while they also establish some further properties,  the measure of the excluded set there  is polynomially small, so as far as establishing positive Lyapunov exponents they don't go beyond \cite{sp}. % Bjerklov \cite{Bjerklov_2005} established positivity of Lyapunov exponents for energies outside a set of small (going to zero as $\lambda \to \infty$) measure for $f \in \mathcal{C}^1(\mathbb{T})$. A similar result was proven by Chan, Goldstein, and Schlag for any $\gamma$-H\"older continuous $f$ on $\mathbb{T}^2$ \cite{ChanGoldsteinSchlag_prepr2006}. %Chan \cite{Chan_2008} proved a perturbative result on positiv
%ity for all energies for typical (in a certain sense) $\mathcal{C}^3$ potentials on $\mathbb{T}$. 

In addition to the above described methods to prove positivity of the Lyapunov exponents, we mention two alternative approaches.

Firstly, for Jacobi operators with $C,V$ given by trigonometric polynomials, combining invariance of the DOS (see Sec. \ref{sec_spectralprop_signatures}) with Thouless' formula (\ref{eq_thouless}) allows to extract lower bounds for the Lyapunov exponents. This strategy has been successfully applied for the AMO and extended Harper's model \cite{MandelshtamZhitomirskaya_1991}, as well as more generally for a Schr\"odinger operator with $V$ given by a linear combination of cosine terms \cite{HaroPuig_2013}. 

Secondly, even though there are no closed expressions for the LE in general, the {\it{Herman-Avila-Bochi formula}} \cite{Herman_1983, AvilaBochi_2002} provides insight for the class of non-uniformly hyperbolic cocycles of the form $(T,AR_\phi)$.  Here, $A: X \to SL(2,\mathbb{R})$ is measurable on a compact space $X$ with $\log\Vert A(.) \Vert \in L^1(X, \ud \mu)$, $R_\phi$ is a rotation by $\phi$, and $T$ is an ergodic transformation on $X$ for some underlying probability measure $\mu$. In this case, the Herman-Avila-Bochi formula asserts that
\begin{equation} \label{eq_HAB}
\int_\mathbb{T} L(T, AR_\phi) \ud \phi = \int_X \log \dfrac{\Vert A(x) \Vert + \Vert A(x) \Vert^{-1}    }{2} \ud \mu(x) ~\mbox{.}
\end{equation}
Zhang used (\ref{eq_HAB}) to prove lower bounds for the Lyapunov exponent of analytic quasi-periodic Schr\"odinger and Szeg\H{o} cocycles \cite{Zhang_2012}. For the Schr\"odinger case the results in \cite{Zhang_2012} recovered Theorem \ref{thm_lyapbourg} for $\nu = 1$.

% To conclude,
We mention that from a more general point of view, Avila proved that for a quasi-periodic Schr\"odinger operator, the set of $V \in \mathcal{C}(\mathbb{T}^\nu)$ giving rise to positive LE for a dense subset of $E \in \mathbb{R}$ is topologically generic. Moreover, topological genericity is shown to be complemented by measure-theoretical prevalence \cite{Avila_2011}.

Finally, the a.e. positivity of the Lyapunov exponents also holds for discontinuous $f,$ for any $\lambda$ (and general continuous dynamics) \cite{damkil}. For monotone $f$ it can even be minorated uniformly for $\lambda>\lambda(f)$ (see Theorem \ref{kach}).

\subsection{Corollaries of positive Lyapunov exponents}\label{sec_supercritical_localization}

Besides absence of absolutely continuous spectrum, guaranteed as a consequence of positive Lyapunov exponents for general ergodic operators,  in the specific situation of quasiperiodic operators, positive Lyapunov exponents have a range of further powerful corollaries, especially for analytic $V.$ The theory has been mostly developed for Schr\"odinger operators, but a few results have been extended to the Jacobi case
(which is usually non-trivial in case singularities are involved). We will comment  on the Jacobi extensions after formulating theorems that combine Schr\"odinger results. Moreover, we will first formulate the results whose proof currently require  analyticity of the potential, and afterwards will list those that are either known to hold more generally or for more specific less regular $V.$
 
%As with the KAM methods,
Positivity of Lyapunov exponents was first used as the main input incorporating all the ``largeness'' of the coupling,  in the proof of localization for the super-critical almost Mathieu operator  \cite{j}. Prior to that localization results have been perturbative, the strongest result being \cite{eli}, for general Gevrey class potentials.

The method in \cite{j} while so far the only one available allowing
precise arithmetic conditions (see Section \ref{sec_models}) also works in certain other situations, e.g. for the extended Harper's model \cite{jks} or monotone potentials \cite{jkach}. However, it does not extend easily to the multi-frequency or even general analytic potentials.

A robust method for the proof of localization for analytic potentials was developed by Bourgain-Goldstein \cite{BourgainGoldstein_2000}
which allowed them to obtain a measure-theoretic version of the
localization result for the general real analytic as well as the
multi-frequency case. Note, that essentially no results were
previously available for the multifrequency case, even perturbative.
\begin{theorem} \label{thbg}
Let $V$ be  non-constant, real analytic on $\mathbb{T}^\nu$ and $H_{\alpha,\theta}$ given by (\ref{eq_schrodingerop}). Suppose that for a full measure set of $\alpha \in \mathbb{T}^\nu$, $L(\alpha, E) > 0$ for all $E \in [E_1, E_2]$. Then, for almost every $\alpha \in \mathbb{T}^\nu$, the spectrum of $H_{\alpha, \theta}$ in $[E_1, E_2]$ is purely point with exponentially decaying eigenfunctions, for almost every $\theta \in \mathbb{T}^\nu$.
%Let $V$ be  non-constant real analytic on $\TT^\nu$ and $H_{\alpha,\theta}$ given by
%(\ref{eq_schrodingerop}). Suppose  $L(E,\alpha)>0$ for all $E\in [E_1,E_2]$ and
%a.e. $\alpha \in \TT^\nu$. Then for any $\theta,$ $H_{\alpha,\theta}$ has Anderson localization in
%$[E_1,E_2]$ for a.e. $\alpha.$ 
\end{theorem}
We mention for completeness that \cite{BourgainGoldstein_2000} in fact establishes that for every {\it{fixed}} $\theta \in \mathbb{T}^\nu$, $H_{\alpha, \theta}$ exhibits spectral localization for almost every $\alpha \in \mathbb{T}^\nu$ (depending in $\theta$), which implies above formulation by Fubini.

Combining this with Theorem \ref{thm_lyapbourg}, Bourgain \cite{Bourgain_2005} obtained that
for $\lambda > \lambda(v),$ $H_{\alpha,\theta}$ as above 
 satisfies Anderson localization for a.e. $\alpha.$ Moreover, as follows then from \cite{bjdyn}, $H_{\alpha,\theta}$ is also dynamically localized.
Theorem \ref{thbg}  extends to potentials belonging to certain Gevrey classes \cite{SKlein_2005}.
One very important ingredient of this method is the theory
of semi-algebraic sets that allows one to obtain polynomial algebraic
complexity bounds for certain ``exceptional'' sets. Combined
with measure estimates coming from the large deviation analysis of 
${1\over n}\ln ||M_n(\theta)||$ (using subharmonic function theory
and involving approximate Lyapunov exponents; see also Sec. \ref{sec_continuity LE}), this theory provides necessary information on the geometric structure of those exceptional
sets. Such algebraic complexity bounds also exist for the almost
Mathieu operator and are actually
sharp  albeit  trivial in
this case due to the specific nature of the cosine.

Let $\frac {p_n} {q_n}$ be the approximants of $\alpha \in \RR \setminus \QQ$. 
Let
\begin{equation}\label{beta}
\beta=\beta(\alpha)=\limsup_{n \to \infty} \frac {\ln q_{n+1}} {q_n}.
\end{equation}

Some further corollaries of positive Lyapunov exponents for analytic sampling functions $V$ and $\nu=1$ are summarized in the following theorem

\begin{theorem} \label{corlyap}

If $V$ is analytic and $L(E,\alpha)>0$ for all $\alpha\in\RR\backslash\QQ$ and all $E \in \mathbb{R}$, then
\begin{enumerate}
\item For any $\theta$ the spectral measure is of full spectral dimension if and only if $\beta(\alpha)=\infty$\cite{jz}
\item For a.e. $\alpha$ 
%satisfying a strong Diophantine condition, 
%$\|q\om\|\ge \frac C {q\ln^Aq}$ for some $C,A,$ 
the spectrum is a 
Cantor set \cite{gs3}

%\item  
\item For $\alpha$ satisfying a strong Diophantine condition, % $\|q\alpha\|\ge \frac C {q\ln^Aq}$ for some $C,A,$ 
the integrated density of states (and the Lyapunov exponent as a function of energy) is H\"older continuous \cite{GoldsteinSchlag_2001} and absolutely continuous \cite{gs2} (see Section \ref{sec_continuity LE})
\item Under the above condition on $\alpha$ the spectrum is homogeneous in the Carleson sense \cite{dgsv}
\item For Diophantine $\alpha$, spectral gaps are almost 
Lipshitz (i.e. a logarithmic correction) continuous in 
frequency \cite{jk}
\item For $\alpha$ satisfying a strong Diophantine condition, there is an almost optimal (up to a logarithmic correction) separation of eigenvalues of finite cutoffs outside sets of small measure  for frequencies outside small sets. \cite{bindervoda}
%\item  Measure of the spectrum is continuous in 
%frequency \ci{jk}  
\end{enumerate}
\end{theorem}

(3),(6) have also been proved for Jacobi matrices \cite{taovoda,bindervoda} and (2) was extended to $d>1$ \cite{DuarteKlein_2013_1}, see Section \ref{sec_continuity LE}.
All the statements can be made local in the energy and 
the frequency in a natural way. Full spectral dimensionality is defined through the the boundary behavior of the Borel transform
of the spectral measure. It implies a range of properties, in particular, packing dimension one and quasiballistic quantum dynamics, see Theorem \ref{jz}.

Some weaker statements are available 
for $\nu>1$ \cite{GoldsteinSchlag_2001} or $V$ belonging to certain Gevrey classes 
\cite{SKlein_2005}. The H\"older exponent can be estimated by $\frac{1}{2k}-\epsilon$ for $f$ being in a small $C^{\infty}$ neighborhood of a 
trigonometric polynomial of degree $k$ \cite{gs2,taovoda}, with an estimate becoming almost precise ($\frac{1}{2} -\epsilon$) 
for the almost Mathieu case (where a precise result is actually available but through the analysis of the dual 
regime, see Theorem \ref{thm_avilajitom_almostred} and note that, by duality, for the almost Mathieu the result follows for all non-critical couplings). % The argument of \ci{gs3} uses some ideas of Sinai \ci{sin} 
% who proved Cantor spectrum 
% for $\cos$-type potentials in the perturbative regime. \ci{gs3} establishes quantitative estimates on separation 
% of eigenvalues using techniques developed in \ci{gs2}. Those techniques extend in turn the ones developed in 
% \ci{gs1}: sharp large deviation estimates for the norms \ci{gs1} and elements \ci{gs2} of the transfer matrix 
% and the avalanche principle. The latter is a general fact about $SL(2,\RR)$ matrices that roughly allows to 
% propagate norm estimates for the large products provided no two nearest neighbors ``almost cancel'' each other.
% This appeared implicitly in \ci{lsy} and was made explicit and developed into a powerful method to study regularity
% properties of the Lyapunov exponents in \ci{gs1}. 

 A Diophantine-type condition is certainly necessary for statements (3),(6) \cite{Bourgain_book_2005}. It is also expected to be necessary for statement (4) and for statement (5) (with only 1/2-H\"older regularity holding in general). It is not entirely clear whether it is necessary for the Cantor spectrum.

While analyticity is important at least for continuity of the Lyapunov exponents, there are a number of corollaries of positive Lyapunov exponents for potentials with much less regularity.

\begin{theorem} \label{thm_Coro_positiveLE_Holder}
 If $V$ is $C^{\gamma}$ and $L(E,\alpha)>0$ for all $\alpha\in\RR\backslash\QQ, E,$ then
\begin{enumerate}
\item 
For all $\theta$ the spectral measures have zero 
Hausdorff dimension (\cite{jlast2})
\item  For $\gamma>1/2,$ the set $S_+$ defined in (\ref{eq_splus}) is continuous in 
frequency \cite{JitomirskayaMavi_preprint_2012}  

\item For  all $\alpha,\theta$ the lower (and for Diophantine $\alpha$ and all $\theta$ the upper)  spreading rate  associated with the slowest spreading portion of the wave packet is equal to zero \cite{jmavi2}
\item If, in addition,  Lyapunov exponents are continuous in energy (in particular, for all analytic $V$)  then for all $\alpha,\theta$ the lower transport exponents are equal to zero \cite{jmavi2}
\item If the Lyapunov exponents are continuous in energy, then for Diophantine $\alpha$ and all $\theta$ the upper transport exponents are equal 
to zero \cite{jmavi2}
\end{enumerate}
\end{theorem}

Here, the upper and lower transport exponents refer to roughly the
upper and lower (in time) power-law growth rates of the moments of
the position operator. Lower exponent being zero corresponds to a very slow
growth along a sequence of scales and upper being zero corresponds to ``almost'' dynamical
localization (as far as power laws of growth are concerned),
coexisting in some cases with singular continuous spectrum. 

The first part was proved in \cite{jlast2} for trigonometric polynomials, but can be extended using the method of \cite{jmavi2} to the $C^\gamma$ setting. The fact that positive Lyapunov exponents imply zero dimensionality for a.e. phase for general ergodic potentials was shown by Simon \cite{s} using potential theory. ``All $\theta$'' is the more delicate part here. For the Diophantine case some results also hold for the shifts of higher-dimensional tori, which is significantly more subtle \cite{jh2}

Some important previous results regarding the second part are \cite{jlast2,jk}. (4) and (5) were originally shown for trigonometric polynomials in \cite{dt}.
The fact that a Diophantine condition is needed in (4)  is illustrated by the following 

\begin{theorem}\label{jz}
 If $V$ is $C^{\gamma}$ then for Liouville $\alpha$ and all $\theta$  the upper transport exponent and the packing dimension of the spectral measure is equal to one.
\end{theorem}

Thus for Liouville $\alpha$ there is a very slow growth at some scales and almost ballistic growth at others. This result is quantitative, providing an estimate on the packing dimension/upper transport exponent dependent on the certain rate or growth of denominators of rational approximants to  $\alpha,$ the quantity $\beta(\alpha),$ (see (\ref{beta})).

\subsubsection {Localization-type results for non-analytic potentials}

The first results on localization for quasiperiodic potentials \cite{Sinai_1987} and \cite{FrlichSpencerWittwer_1990}  were KAM type. They were aimed at the almost Mathieu operator but did not use the analyticity, focusing instead on the $\cos$-type structure. The required smoothness is then only $C^2$ with results being perturbative but otherwise resembling those for the analytic case. Recently this model was approached by purely dynamical methods in \cite{wz1,wz2}. We summarize the results that do not require energy exclusion in
\begin{theorem}Assume $\alpha$ is Diophantine and $V = \lambda f$ such that $f \in C^2$ and has exactly two non-degenerate extrema. Then for $\lambda>\lambda(\alpha,f)$:
\begin{enumerate}
\item for all $E,$ Lyapunov exponents $L(E)>(1-\epsilon)\log \lambda.$ \cite{wz1}
\item the Lyapunov exponent is continuous in energy; moreover, the Lyapunov exponent and the integrated density of states are at least sub-H\"older continuous \cite{wz1}
\item the spectrum is a Cantor set \cite{Sinai_1987,wz2}
\item for a.e. such $\alpha$ there is localization for a.e. $\theta.$\cite{Sinai_1987,FrlichSpencerWittwer_1990}.
\end{enumerate}
\end{theorem}

In this case none of the proofs exploit positivity of the Lyapunov exponents, establishing instead certain technical statements from which both such positivity and the other results follow. The method of \cite{wz1,wz2} is based on the careful analysis of stable and unstable directions and inductive classification of their intersections in the associated cocycle dynamics. It is further developing the ideas of \cite{young}, see also Sec. \ref{sec_continuity LE}. It should be noted that the proof of Cantor spectrum in \cite{gs3} uses the ``follow the resonances''  idea of Sinai's approach \cite{Sinai_1987}. The methods of \cite{Sinai_1987,FrlichSpencerWittwer_1990} have proved to be difficult to generalize to non-$\cos$-type functions. It is not yet clear whether the method of \cite{wz1,wz2} can be extended with less difficulties.

As resonances both lead to the Cantor spectrum and create obstacles for localization, the latter should be more manageable when the resonances are minimized. This is in the foundation of

\begin{theorem}\cite{jkach}\label{kach}
Assume $V=\lambda f$ is a monotone Lipshitz function with a Lipshitz inverse.  Then for Diophantine $\alpha$,
\begin{enumerate}
\item The integrated density of states is Lipshitz continuous
\item The Lyapunov exponent is uniformly bounded from below for all $E$ for $\lambda>\lambda(f)$ (depending only on the Lipshitz constants)
\item Positive Lyapunov exponent on a set of nonzero spectral measure implies Anderson localization on this set. 
\item The corresponding eigenfunctions are uniformly localized on a set where the Lyapunov exponent is uniformly bounded from below.
\end{enumerate}
\end{theorem}

A Diophantine condition is clearly necessary, since for Liouville $\alpha$ a Gordon-type argument leads to singular-continuous spectrum. The proof of (3),(4) uses the strategy of \cite{j}, which is actually somewhat simplified in this case. Theorem \ref{kach} leads to the following

\begin{coro}\cite{jkach}\label{cor}
For all $\alpha$ and $V= \lambda f$ as in Theorem \ref{kach}, 
\begin{enumerate}
\item For $\lambda>\lambda(f)$ and all $\theta$ operator $H_{\lambda; \alpha,\theta}$ exhibits uniform spectral and uniform dynamical localization
\item For all $\lambda>0$ and a.e. $\theta$ $H_{\lambda; \alpha, \theta}$ has pure point spectrum with exponentially decaying eigenfunctions.
\end{enumerate}
\end{coro}

Therefore, besides analytic $V$, localization has been established, for $C^2$ $\cos$-type potentials (perturbatively) and, for all monotone potentials with a bi-Lipshitz condition,  for {\it all} non-zero couplings. 

\section{Avila's global theory} \label{sec_global}

In his ``{\it{Global theory of one-frequency Schr\"odinger operators}}'' \cite{Avila_globalthy_published}, A. Avila answers the question whether the spectral properties of the AMO, with its key-feature, appearance of a ``critical phase ($\lambda=1$),''  at the transition point between localization (``insulator-like phase,'' $\lambda>1$) and purely ac spectrum (``metallic phase,'' $\lambda<1$), is typical among quasi-periodic Schr\"odinger operators  with $V \in \mathcal{C}^\omega(\mathbb{T}; \mathbb{R})$ and $d=\nu=1$.

To this end, Avila decomposes  the spectrum into supercritical ($\Sigma_+$), subcritical ($\Sigma_{sub}$), and critical ($\Sigma_{crit}$) energies, which provides a suitable dynamical framework which generalizes the ``insulator, metallic, and critical phase'' exhibited by the AMO. As outlined in Sec. \ref{sec_parametercomplexification}, the three regimes are quantified by the (phase)-complexified LE, (\ref{eq_defcomplexle}). In particular, $\Sigma_{crit}$ explicitly identifies the contributions from {\it{purely singular}} spectrum in the set of zero Lyapunov exponent (see Theorem \ref{thm_AFK}). %, and is even conjectured to identify purely sc continuous spectrum (CEC, see Conjecture \ref{conj_cec}).

While it was precisely the presence of $\Sigma_{crit}$ which attracted vast interest in the AMO, Avila's global theory shows that the behavior exhibited by the AMO is radically atypical:
\begin{theorem}[``Prevalence of acriticality,'' \cite{Avila_globalthy_published}] \label{thm_Avila_acriticality}
Let $\alpha \in \mathbb{R}\setminus \mathbb{Q}$. For prevalent $V \in \mathcal{C}^\omega(\mathbb{T}; \mathbb{R})$ one has $\Sigma_{crit} = \emptyset$ and there exist $a_1 < b_1 < a_2 < b_2 < \dots < a_n < b_n$ with $\Sigma \subseteq \cup_{j=1}^{n} [a_j,b_j]$ such that energies alternate between subcritical and supercritical behavior along the sequence $\{ \Sigma \cap [a_j, b_j] \}_{1 \leq j \leq n}$.
\end{theorem}
Here, prevalence is a notion of {\it measure-theoretical} typicality in $\mathcal{C}^\omega(\mathbb{T}; \mathbb{R})$: Fixing an exponentially decaying function $\epsilon: \mathbb{N} \to \mathbb{R}_+$, a property is prevalent if it holds a.e. with respect to all shifts of the compactly supported measure $\mu_\epsilon$ obtained by push-forward of the normalized Lebesgue product measure on $\mathbb{D}^\mathbb{N}$ under the embedding, $(t_n) \mapsto \mathrm{Re} \{ \sum_{n \in \mathbb{N}} \epsilon(n) t_n \mathrm{e}^{2 \pi i n x} \} \in \mathcal{C}^\omega(\mathbb{T}; \mathbb{R})$.

The technical key ingredients in the proof of Theorem \ref{thm_Avila_acriticality}, both established in \cite{Avila_globalthy_published} are ``quantization of the acceleration'' (see Sec. \ref{sec_parametercomplexification_phasecomplexfication}) and Theorem \ref{thm_dsAJS} in its original form for $SL(2, \mathbb{C})$ cocycles, the latter of which allows to detect $\mathcal{UH}$ for positive LE through regularity of the cocycle. Immediate important corollaries are:
\begin{itemize}
\item[(Gi)] $\Sigma_{sub}$ is open (in $\Sigma$), a consequence of both quantization and upper semicontinuity of the acceleration; this complements openness of $\Sigma_+$ implied by continuity of the LE. 
\item[(Gii)] More generally, acritical behavior is stable with respect to perturbations in $\alpha$ and $V$; $\Sigma_+$ and $\Sigma_{sub}$ depend continuously in the Hausdorff metric on the perturbations.
\item[(Giii)]  {\it A dynamical dichotomy}: for all $E \in \mathbb{R}$, $\exists y_0 >0$ such that for all $0< \vert y \vert \leq y_0$ either $L(E; y) = 0$  or $(\alpha, A_y^E) \in \mathcal{UH}$ ~.
\end{itemize}

The spectrum can thus be stratified into nested closed sets with strata corresponding to values of {\it{fixed acceleration,}} $\omega(\alpha, A_{y=0}^E) = n \in \mathbb{N} \cup \{ 0 \}$. More generally, allowing for a variation of both the energy and the potential, the same stratification exhibits $\mathcal{C}^\omega$-dependence of the LE when restricted to the strata:
\begin{theorem}[\cite{Avila_globalthy_published}] \label{thm_Avila_strateification}
Let $\alpha \in \mathbb{R} \setminus \mathbb{Q}$ and $V_\lambda$ in $\mathcal{C}^\omega(\mathbb{T}; \mathbb{R})$ be a real analytic parameter family with $\lambda$ ranging in a real analytic manifold $\Lambda$. Then, $L(\alpha, A^{E;\lambda})$ is a $\mathcal{C}^\omega$-stratified function of both $\lambda$ and $E$.
\end{theorem}
This $\mathcal{C}^\omega$-dependence of the LE is  a consequence of the dynamical dichotomy in (Giii): For $(E,\lambda)$ such that $\omega(\alpha, A_{y=0}^{E;\lambda}) = n > 0$ and $y_0 > 0$ as in (Giii), one has 
\begin{equation} \label{eq_global_auxil}
L(\alpha, A^{E;\lambda}) = L(\alpha, A_{y_0}^{E;\lambda}) -  2 \pi y_0 n ~\mbox{,}
\end{equation}
which, since $(\alpha, A_{y_0}^{E;\lambda}) \in \mathcal{UH}$, depends analytically on $(E,\lambda)$. Thus, if $(E,\lambda)$ is critical with $\omega(\alpha, A_{y=0}^{E;\lambda}) = n$ (``{\it{$E$ is critical of degree $n$}}''), then $(E,\lambda)$ is contained in the zero set of a $\mathcal{C}^\omega$ function, given by the right-hand side of (\ref{eq_global_auxil}). Avila further shows that each of these countably many $\mathcal{C}^\omega$-functions are {\it{submersions}} locally about critical behavior. This leads to the following structural characterization of the locus of criticality:
\begin{theorem}[\cite{Avila_2009}] \label{thm_Avila_locusOfCriticality}
Let $\alpha \in \mathbb{R}\setminus \mathbb{Q}$ and $\delta>0$. The set of potentials and energies $(V,E)$ such that $E$ is critical is contained in a countable union of codimension-one analytic submanifolds of $\mathcal{C}_\delta^\omega(\mathbb{T}; \mathbb{R}) \times \mathbb{R}$.
\end{theorem}

While Theorem \ref{thm_Avila_locusOfCriticality} already implies that for prevalent $V$, $\Sigma_{crit}$ can be at most {\it{countable}}, \cite{Avila_globalthy_published} subsequently goes on to prove that given $V \in \mathcal{C}^\omega(\mathbb{T}; \mathbb{R})$, criticality with energies of degree at most $n$ can be ``destroyed'' by a small typical perturbation by trigonometric polynomials of sufficiently large degree. These perturbations yield critical energies of degree at most $n-1$, which, upon iteration, yields Theorem \ref{thm_Avila_acriticality}.

\section{Explicit models} \label{sec_models}
It is fair to say that the development of the spectral theory of quasiperiodic operators have been largely centered around and driven by several explicit models, all coming from physics.

In this section we will present the highlights of the current state-of-the-art for the following three models: almost Mathieu operator, extended Harper's and the Maryland model. (Another popular model, the Fibonacci operator, is described in detail in \cite{Damanik_review2014}). Even though the almost Mathieu operator is a particular case of the extended Harper's model, the corresponding results are often more detailed/complete so we present them separately.  Those models all demonstrate interesting dependence on the arithmetics of parameters (often even when the final results turn out to be independent of such arithmetics the proofs have to be different in different regions) and have traditionally been approached in a perturbative way: through KAM-type schemes in the regime of large couplings/ reducibility (which, being a KAM-type scheme, only works for Diophantine frequencies) or through perturbation of periodic operators (Liouville frequency). Even when the KAM arguments have been replaced by the non-perturbative ones allowing to treat more or even all couplings, frequencies that are neither far from nor close enough to rationals presented a challenge as for them there was nothing left to perturb about.  A remarkable relatively recent development concerning the explicit models is that very precise results have become possible: not only many facts have been established for a.e. frequencies and phases, but in many cases it has become possible to go deeper in the arithmetics and either establish precise arithmetic transitions or even obtain results for all values of parameters. 

\subsection {The almost Mathieu operator}%Operator $H_{\lambda,\alpha,\theta}$ given by (\ref{eq_schrodinger_op}) with $V=2\lambda\cos 2\pi \theta$ iscalled the almost Mathieu operator.  It is
The AMO ($V (\theta) = 2 \lambda \cos(2 \pi \theta)$) is the central quasi-periodic model mainly due to being the first of its kind that is coming from physics and attracting continued interest there. It first appeared in the work of Peierls \cite {Peierls_1933} and arises as related, in two different ways, to a two-dimensional
electron subject to a perpendicular magnetic field \cite{Harper_1955,rauh}, see Section \ref{sec_models_physics}. It plays
a central role in the Thouless et al theory of the integer quantum Hall
effect \cite{ThoulessKohmotoNightingaleNijs_1982}. Also, even though the simplest possible analytic potential, it seems to represent most of the
nontrivial properties expected to be encountered in the more general
case. On the other hand it has a very special feature: the duality
transform (\ref{eq_dualop_jacobi}) maps $H_{\lambda; \alpha, \theta}$ to $H_{1/
\lambda; \alpha, \theta},$ hence $\lambda=1$ is the self-dual (also called critical) point. The development of the 
rigorous theory of this model (and of general quasiperiodic operators along
with it) was strongly motivated by the numerical study of Hofstadter
in 1976,
the famous Hofstadter's butterfly \cite{Hofstadter_1976}, and was guided for a 
long time by two conjectures formulated by Aubry and Andr\'e in \cite{AubryAndre_1980} and heavily popularized 
in several of B. Simon's articles in the early 80s. The more detailed history up to 2007 is presented in \cite{Jitomirskaya_review_2007,lastrev}. Here we present the situation with the current knowledge, giving slightly more detail concerning results since 2007. While a number of statements of previously formulated more general theorems apply to this operator family, we concentrate here only on the ones that are (currently) almost Mathieu specific. 
We will start with the results where the arithmetic nature of $\alpha$ does not play a role in the formulation (although it does, in rather remarkable ways, in the proofs). For statements with a long history we include the references to the most major (not all!) partial results, with the work where a corresponding theorem was proved as stated listed last. 
\begin{theorem}\label{amgeneral} For all irrational $\alpha$ we have
\begin{enumerate}
\item The measure of the spectrum of $H_{\lambda; \alpha,\theta}$ is equal to $|4-2|\lambda||$ for all $\theta,\lambda$ \cite{l1,l2,jk,AvilaKrikorian_2006}.
\item For all $\lambda$ and all $E$ in the spectrum of $H_{\lambda; \alpha,\theta}$ the Lyapunov exponent is equal to $\max \{\ln |\lambda|,0\}$ \cite{BourgainJitomirskaya_2002}.
\item For all $\lambda\not= 0,$ the spectrum of $H_{\lambda; \alpha,\theta}$ is a Cantor set. \cite{puig, AJ2}
\item For all $\lambda <1$ (the sub-critical case) and all $\theta$ the spectrum is absolutely continuous. \cite{l1,j,Avila_2008}.  

\end{enumerate}
\end{theorem}
It is remarkable that with prior work establishing (1),(3),(4) for a.e. $\alpha$ (and/or $\theta$), the final challenge remained to settle the cases of the remaining measure zero set of parameters. Those were formulated as correspondingly problems 5, 4, and 6 in \cite{Simon_2000} and resolved, each in a rather intricate way, in, correspondingly, \cite{AvilaKrikorian_2006,AJ2,Avila_2008}. Additional partial advances on part (4) include \cite{ad,AvilaJitomirskaya_2010}

For the supercritical case, $\lambda>1,$ the arithmetic nature of the frequency $\alpha$ and phase $\theta$ start playing a crucial role not only for the proofs but also for the results.
Recall that $\beta(\alpha)$ is given by (\ref{beta}). For any $\alpha,\theta$ we also define $\delta(\alpha,\theta)\in [0,\infty]$ as
\begin{equation}\label{G.delta}
\delta=\delta(\alpha,\theta)=\limsup_{n \to \infty} \dfrac{-\ln \vert\vert\vert 2 \theta+n \alpha \vert\vert\vert }{\vert n \vert} ~\mbox{.}
\end{equation}

%\newpage
We have
\begin{theorem}\label{amsuper}
\begin{enumerate}
\item For any $\alpha$ and  $1<\lambda < e^{\beta},$ the spectrum is purely singular continuous, for all $\theta.$ \cite{ayz1}
\item For any $\alpha$ and $\lambda > e^{\beta},$ the spectrum is pure point for a.e. $\theta$ \cite{ayz1}
%$\alpha$-Diophantine $\theta$ \cite{jl1}
\item For any $\theta$ and $1<\lambda < e^{\delta},$ the spectrum is purely singular continuous, for all $\alpha.$\cite{jl2}
\item For any $\theta$ and $\lambda > e^{\delta},$ the spectrum is pure point for a.e. $\alpha$ \cite{jl2}
\item  For $\lambda> e^{\beta},$  the spectrum is singular continuous for a certain (arithmetically defined) dense $G_{\delta}$ of $\theta.$
 \cite{JitomirskayaSimon_1994}
\end{enumerate}
\end{theorem}

The fact that for a.e. $\theta$ the spectral transition in frequency happens at $\lambda=e^{\beta}$ and for a.e. $\alpha$ at $\lambda=e^{\delta},$ was first conjectured in \cite{jalmost}. Part (1), for sufficiently large $\beta,$ was first established in \cite{Gordon_1976,AvronSimon_1983}. Part (2), for $\beta=0$ was proved in \cite{j}. %where a.e. $\theta$ is described 
%through an explicit arithmetic condition, so called {\it non-resonance} property, in particular this set contains $0.$ 
This was extended to $\beta<9/16\log\lambda$ in \cite{AJ2} and that argument was pushed to its technical limit, leading to the result for $\beta<2/3\log\lambda,$ in \cite{liu}. The existence of a dense collection of eigenvalues was proved in \cite{yz} for $\beta < \log\lambda - C.$   The final localization result \cite{ayz1} does not use a direct localization argument  but relies instead on the almost reducibility theorem \cite{Avila_prep_ARC_1} and duality, leading to the sharp statement (but losing the precise control on the phases).  See also \cite{jitkach} where a simple alternative way to
argue completeness in the duality argument was presented. However, it turns out that a.e. in (2) and (4) can be described in an arithmetic way. For any $\alpha$, we say that  phase $\theta\in (0,1)$   is
 Diophantine  with respect to $\alpha$ (or $\alpha$-Diophantine) if there exist $ \kappa>0$ and $\nu>0$ such that
 \begin{equation}\label{DCtheta}
   \vert\vert\vert 2\theta-k\alpha \vert\vert\vert > \frac{\kappa}{|k|^{\nu}},
 \end{equation}
 for any $k\in \mathbb{Z} \backslash \{0\}$.
 
 Clearly, for any irrational number $\alpha$, the set of phases which
 are  Diophantine with respect to $\alpha$ is of full Lebesgue measure. Then
\begin{theorem}\label{amsuper2}
\begin{enumerate}
%\item For any $\alpha$ and  $1<\lambda < e^{\beta}$ the spectrum is purely singular continuous, for all $\theta.$ \cite{ayz1}
\item For any $\alpha$ and $\lambda > e^{\beta}$ the spectrum is pure point for $\alpha$-Diophantine $\theta$ \cite{jl1}
%$\alpha$-Diophantine $\theta$ \cite{jl1}
%\item For any $\theta$ and $1<\lambda < e^{\delta}$ the spectrum is purely singular continuous, for all $\alpha.$\cite{jl2}
\item For any $\theta$ and $\lambda > e^{\delta}$ the spectrum is pure point for Diophantine $\alpha$ \cite{jl2}

\end{enumerate}
\end{theorem}
We note that the proof of (1) in \cite{jl1} extends the argument of \cite{AJ2} adding an effective new idea on how to treat the resonant points. 
Thus there are sharp criteria that describe the spectral transition in both phase and frequency.

% Then
% \begin{theorem}\label{amsuperphase} \cite{jl2} For any Diophantine $\alpha$
% \begin{enumerate}
% \item For $1<\lambda < e^{\delta}$ the spectrum is purely singular continuous 
% \item For $\lambda > e^{\delta}$ the spectrum is pure point 

% \end{enumerate}
% \end{theorem}
In both \cite{jl1} and \cite{jl2} the proof is constructive and in fact provides also sharp lower bounds, thus a {\it precise} exponential asymptotics of the eigenfunctions and transfer-matrices. 
We say that $\phi$ is a generalized  eigenfunction of $H$ with generalized
eigenvalue $E$, if
   \begin{equation} \label{shn}
     H\phi=E\phi  ,\text{ and }  |\phi(k)|\leq C(1+|k|).
  \end{equation}
For a generalized eigenfunction $\phi$ let
$ U(k) =\left(\begin{array}{c}
        \phi(k)\\
       \phi({k-1})
     \end{array}\right)
 $.
\begin{theorem}\label{Maintheoremdecay}\cite{jl1,jl2} In the entire regime of Theorem \ref{amsuper}, (2) or Theorem \ref{amsuper2}, (2), there exist explicit functions $f(n),g(n)$ depending only on $\alpha$ in the former and on $\alpha,\theta$ in the latter case, so that if
$E$ is a generalized eigenvalue of $H_{\lambda,\alpha,\theta}$ and  $\phi$ is
the  generalized eigenfunction, %and   $ \psi$ be any other linear independent solution respected to  $\phi$.
then for any $\varepsilon>0$, there exists $K$ such that for any $|k|\geq K$,  $U(k)$ and $A_k$ %and $ \tilde{U}(k)$
satisfy
 %Let  $\frac{q_n}{3}\leq|k|<\frac{q_{n+1}}{3}$.
\begin{equation}\label{G.Asymptotics}
 f(|k|)e^{-\varepsilon|k|} \leq ||U(k)||\leq f(|k|)e^{\varepsilon|k|},
\end{equation}
and
\begin{equation}\label{G.Asymptoticstransfer}
 g(|k|)e^{-\varepsilon|k|} \leq ||A_k||\leq g(|k|)e^{\varepsilon|k|}.
\end{equation}
\end{theorem}
Therefore it provides a precise description of the non-Lyapunov behavior in this case, leading to some surprising phenomena such as e.g. exponentially strong tangencies between expanding and contracting directions at resonant points. The generalized eigenfunctions in the singular continuous regime can also be analyzed, leading to nontrivial bounds on quantum dynamics and, for example,
\begin{theorem}\label{ther}\cite{jlt}
\begin{itemize}
\item if $|\lambda|=e^{\beta(\alpha)}$, then all  spectral measures  are zero packing dimensional .
 \item if $e^{\frac{1}{2}\beta(\alpha)}< |\lambda|< e^{\beta(\alpha)}$, then  for any $\varepsilon>0$,
all  spectral measurs are $2(1-\frac{\ln|\lambda|}{\beta(\alpha)})+\varepsilon$ packing singular.
\end{itemize}
\end{theorem}
This should be contrasted with the fact that spectral measures are $(1-\frac{C\ln|\lambda|}{\beta(\alpha)})-\varepsilon$ packing continuous \cite{jz} and of zero Hausdorff dimension \cite{s,jlast2}.

The full arithmetic spectral transition conjecture \cite{jalmost} is that for all irrational $\alpha$ and all $\theta$ the transition happens at $\lambda=e^{\beta+\delta}.$ This is currently still out of reach. It should be noted this is likely specific to the almost Mathieu operator, while (1),(2) of Theorem \ref{amsuper}  and even (1) of Theorem \ref{amsuper2} are expected to be much more universal if $\log\lambda$ is replaced by the Lyapunov exponent. 

Theorems \ref{amsuper},\ref{amsuper2} still leave the values of $\alpha$ with $\lambda=e^{\beta}$ undescribed. It is expected that for them the answer will depend on finer properties of the approximants. In particular,  some such $\alpha$'s have  pure point spectrum, while some others purely singular continuous \cite{ajz,jqz}.

Almost Mathieu operators also exhibit very strong dynamical localization in the supercritical regime 
\begin{theorem} For $\lambda>1$ and $\alpha$ with $\beta(\alpha)$ sufficiently small there is exponential dynamical localization  (or the expected overlap decays exponentially (in space)) \cite{jkru} . Moreover, for Diophantine $\alpha$ the coefficient is precisely equal to $\log\lambda.$ \cite{jkl}
\end{theorem}

A similar property (even just the existence of the exponent, not even speaking of a precise computation) has not yet been established for other quasiperiodic models (aside from the monotone potentials).

Another question on which a recent progress has been made is the {\it{Dry Ten Martini conjecture}}, which says that for all $\lambda$ and all irrational $\alpha$ there is an open gap at each energy with the integrated density of states $N(\alpha,E)\in \ZZ+\alpha\ZZ \mod 1.$ (those are the only possible gaps according to the gap labelling theorem, see Sec. \ref{sec_reducibility}). The word "dry" refers to the fact that it is strictly stronger than proving that the gaps are dense in the spectrum which is known as the Ten Martini Conjecture (see Theorem \ref{amgeneral}, part 2). The martinis offer was made by Mark Kac and popularized by Barry Simon (the original non-dry conjecture is due to Mark Azbel).  The Dry Ten Martini was previously established for Liouville $\alpha$ \cite{cey} and for Diophantine $\alpha$ and non-critical $\lambda$ \cite{AvilaJitomirskaya_2010}, leaving again the gap in the methods between Diophantine and Liouville. Establishing it for {\it all} vs a.e. $\alpha$ and all $\lambda$  would prove the visually obvious fact that the Hofstadter butterfly has wings, which would lead to various interesting corollaries, particularly related to the Quantum Hall effect. A recent result is
\begin{theorem}\cite{ayz2}\label{dry}
For all irrational $\alpha$ and $\lambda\not=0$ the Dry Ten Martini Conjecture holds, except for possibly the case of $|\lambda|=1$ and $\beta(\alpha)=0.$
\end{theorem}
%It remains an open problem to remove the above exceptions.

The spectral theory of the critical case $\lambda=1$ has been described for all but countably many values of parameters. First, it follows from (1) of Theorem \ref{amgeneral} that there is no absolutely continuous spectrum for any $\alpha,\theta,$ thus the question boils down to the absence of eigenvalues.
\begin{theorem}\label{amcrit}
If $\alpha$ is irrational and $\theta\notin \ZZ \alpha + \ZZ$ the spectrum of $H_{1,\alpha,\theta}$ is purely singular continuous \cite{Avila_preprint_2008_2,AvilaJitomirskayaMarx_preprint_2014}.
\end{theorem}
Moreover, the eigenvalues, if  any, cannot belong to $\ell^1$ \cite{delyon}. For $\alpha$ that are well approximated by the rationals absence of eigenvalues is known for all $\theta$ (see Theorem \ref{amsuper}).

Despite so many studies and the well recognized importance of the critical almost Mathieu operator it is still not very well understood, with most results proved either indirectly or by approximation from the non-critical cases. In particular, the open problems for this case  include
\begin{enumerate}
\item The Dry Ten Martini problem
\item Spectral decomposition for $\alpha$-rational phases
\item Hausdorff (or other) dimension of the spectrum/spectral measures 
\end{enumerate}
All these questions are currently unsolved for a.e. $\alpha$. In particular,  Hausdorff dimension has only been computed for a measure zero set of parameters \cite{l2,lastshamis,mirathesis}

\subsection {The extended Harper model} 
The {\it{extended Harper's model}}  is a model from solid state physics defined by the operator $H_{\lambda; \alpha, \theta}$ given by (\ref{eq_operator_def}) with 
\begin{equation} \label{eq_hamiltonian1}
C(\theta)= C_{\lambda}(\theta) := \lambda_{1} \mathrm{e}^{-2\pi i (\theta+\frac{\alpha}{2})} + \lambda_{2} + \lambda_{3} \mathrm{e}^{2 \pi i (\theta+\frac{\alpha}{2})} ~\mbox{,}
~ V(\theta)  := 2 \cos(2 \pi \theta) ~\mbox{,}
\end{equation}
where $\lambda=(\lambda_1, \lambda_2, \lambda_3) \in \mathbb{R}^3$ denotes the coupling parameter of the model. % following quasi-periodic Jacobi operator acting on $\mathit{l}^2(\mathbb{Z})$,
% \begin{eqnarray} \label{eq_hamiltonian}
% & (H_{\theta;\lambda, \alpha} \psi)_k := v(\theta + \alpha k) \psi_{k} + c_{\lambda}(\theta + \alpha k) \psi_{k+1} + \overline{c_{\lambda}(\theta + \alpha (k-1))} \psi_{k-1} ~\mbox{.}
% \end{eqnarray}
%Here, $\alpha$ is a fixed irrational, $\theta$ varies in $\mathbb{T}:=\mathbb{R}/\mathbb{Z}$, and

% Physically, extended Harper's model describes the influence of a transversal magnetic field of flux $\alpha$ on a single tight-binding electron in a 2-dimensional crystal layer, allowing for various lattice geometries. 
It arises in physics as described in Section \ref{sec_models_physics}.
The almost Mathieu operator is a particular case with $ \lambda_{1}= \lambda_{3}=0.$ The extended Harper's model was originally proposed by D. J. Thouless in 1983 in context with the integer quantum Hall effect \cite{Thouless_1983}. As for certain physically relevant values of $\lambda$ the corresponding Jacobi operator is {\it{singular}}, this model has been the main motivator in the development of the general theory of (singular)  Jacobi  operators. 

The parameter space of the extended Harpers model is naturally partitioned into the three regions I-III depicted in Fig. 1, with regions I and II being dual to each other and the union of region III and line L2 being self-dual (and most difficult to treat, with controversies even in Physics literature).

%\newpage
\begin{figure}[ht]  \label{figure_1}
\includegraphics[width=0.5\textwidth]{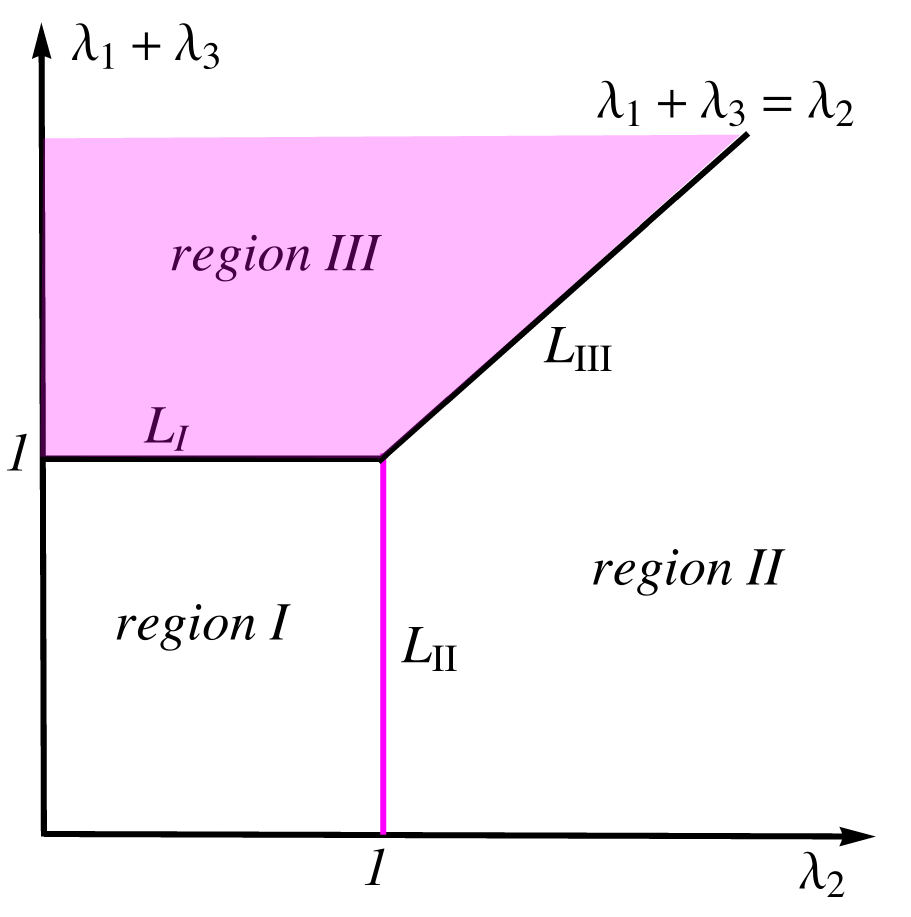}
\caption{Partitioning of the space of coupling constants $\lambda=(\lambda_1,\lambda_2,\lambda_3)$ for extended Harper's model. The interesting self-dual regime is colored in red.}
\end{figure}

The Lyapunov exponents of this model have been determined exactly for {\it all} values of parameters:
\begin{theorem}\label{ehm}\cite{JitomirskayaMarx_2012}
Fix an irrational frequency $\alpha$. Then the Lyapunov exponent on the spectrum is zero within both region II and III. In region I it is given by the formula,
\begin{equation} \label{eq_thoulessform}
\begin{cases}\log \left(  \dfrac{1+\sqrt{1 - 4\lambda_{1} \lambda_{3}}}{2 \lambda_{1}}\right) & \mbox{, if} ~\lambda_{1} \geq \lambda_{3}, ~\lambda_{2} \leq \lambda_{3} + \lambda_{1} ~\mbox{,}  \\
\log \left(  \dfrac{1+\sqrt{1 - 4\lambda_{1} \lambda_{3}}}{2 \lambda_{3}}\right) & \mbox{, if} ~\lambda_{3} \geq \lambda_{1}, ~\lambda_{2} \leq \lambda_{3} + \lambda_{1} ~\mbox{,}  \\
\log \left(  \dfrac{1+\sqrt{1 - 4\lambda_{1} \lambda_{3}}}{\lambda_{2} + \sqrt{\lambda_{2}^{2} - 4 \lambda_{1} \lambda_{3}}}\right) & ~\mbox{, if} ~\lambda_{2} \geq \lambda_{3} + \lambda_{1} ~\mbox{.} 
\end{cases}
\end{equation}
\end{theorem}
Computation of the LE for extended Harper's model was possible using complexification of the phase together with the analytic properties of the complexified LE, $L(E;y)$, see (\ref{eq_defcomplexle}). The strategy dubbed ``{\it{method of almost constant cocycles}},'' relies on the observation that the Jacobi cocycles in the limit $y \to \infty$ are close to a constant cocycle inducing a $\mathcal{DS}$. Since this method allowed to compute the complexified LE, it immediately implied a complete description of the spectral properties in the sense of Avila's global theory. We mention that these ideas can be used more generally for Jacobi operators with sampling functions given by trigonometric polynomials, e.g. \cite{Marx_Shou_Wellens_2015} employs it to obtain sufficient criteria for sub-critical behavior for a generalized almost Mathieu operator with potential given by a linear combination of cosines. 

Like for the almost Mathieu operator, the spectral theory can now be described fully for all $\lambda$ and  almost all $\alpha,\theta:$
\begin{theorem} \label{thm_ehmspectral}
%Assuming the almost reducibility conjecture, 
For all Diophantine $\alpha$ for Region I, and for all irrational $\alpha$ for regions II, III, Fig. 2
%\ref{figure_5} 
represents the Lebesgue decomposition of the spectrum of $H_{\theta;\lambda, \alpha}$ for a.e. $\theta$.
\begin{figure}[ht] \label{figure_5}
\centering
\subfigure[$\lambda_1 \neq \lambda_3$]{
\includegraphics[width=0.4\textwidth]{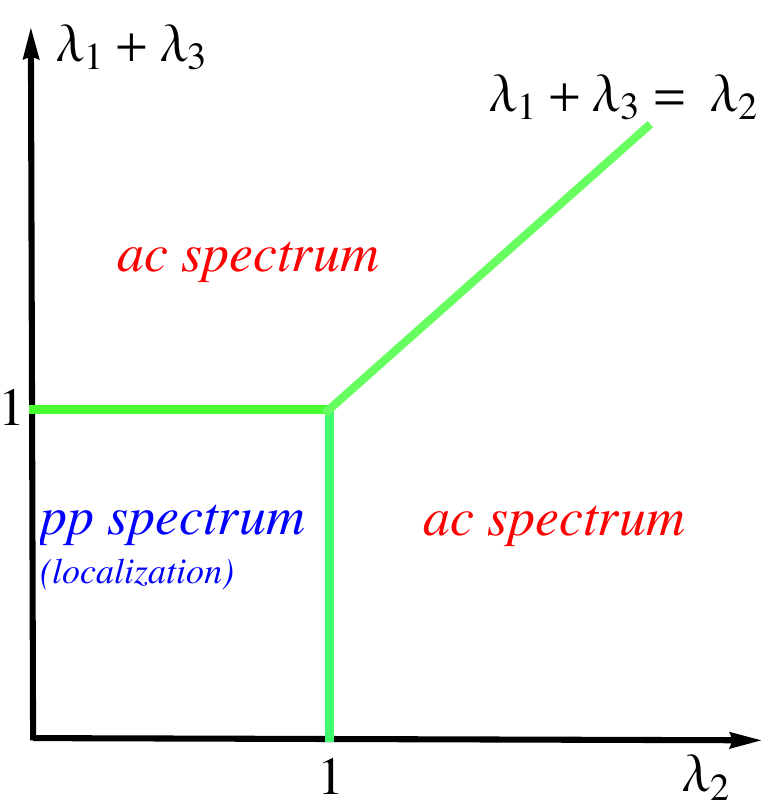} \label{figure_3}
}
\subfigure[$\lambda_1 = \lambda_3$]{
\includegraphics[width=0.4\textwidth]{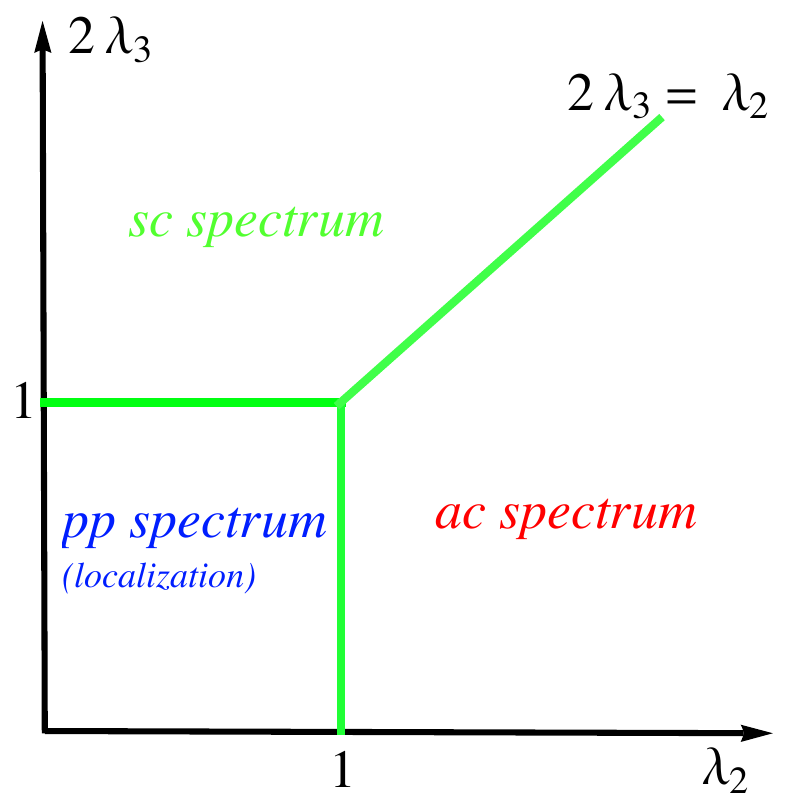} \label{figure_4}
}
\caption{Spectral theory of extended Harper's model. Green indicates (purely) singular continuous spectrum. The spectral properties of extended Harper's model crucially depend on the symmetry of NNN interaction. Particularly noteworthy is the collapse in the self-dual regime, from purely absolutely continuous spectrum for $\lambda_1 \neq \lambda_3$ to purely singular continuous spectrum once $\lambda_1 = \lambda_3$. Anderson localization in region I had been proven before in \cite{jks}.} 
\end{figure}
\end{theorem}
The result for the region I is due to \cite{jks}, and for the region II,III to \cite{jmarx, AvilaJitomirskayaMarx_preprint_2014}. For Liouville $\alpha$ in region I the spectrum is singular continuous \cite{Koslover_2005}, but the precise arithmetic transition, like for the almost Mathieu operator, has not yet been explored. 

As for the measure of the spectrum, it has been determined as of this writing in the non-self-dual region only. More precisely, let $a\leq b\leq c$ be the ordering of $1,\lambda_2,\lambda_1+\lambda_3.$ We have 

\begin{theorem}\cite{rui1} For almost all $alpha,$ measure of the spectrum in regions I,II is equal to $4(c-b).$

% \begin{itemize}
% \item  in the region $I\cap\{\lambda_1+\lambda_3\leq \lambda_2\}\cup II\cap \{\lambda_1+\lambda_3\leq 1\}$ measure of the spectrum is equal to $|4|\lambda_2-1|.$
% \item  in the region $I\cap\{\lambda_1+\lambda_3>\lambda_2\}$ measure of the spectrum is equal to $4|1-(\lambda_1+\lambda_3)|$.
% \item   in the region $ II\cap \{\lambda_1+\lambda_3 > 1\}$ measure of the spectrum is equal to $4|\lambda_2-(\lambda_1+\lambda_3)|$.
% \end{itemize}
\end{theorem}

The Cantor nature of the spectrum has been determined for Diophantine $\alpha$ in regions I, II, moreover:
\begin{theorem}\cite{rui2}  For any $\lambda\in I, II,$ if $\beta(\alpha)=0,$ the dry ten martini holds.
\end{theorem}

 The situation so far proves difficult for the remaining parameters.

\subsection{The Maryland model}

The Maryland model
is given by
 \begin{equation}\label{MME}
(h_{\lambda,\alpha,\theta}u) _n=u_{n+1}+u_{n-1}+  \lambda \tan \pi(\theta+n\alpha)u_n.
 \end{equation}
%with $ \theta\notin  1/2 + \alpha\mathbb{Z}+\mathbb{Z}$ (so the potential is well defined).
It was proposed by Grempel, Fishman, and Prange
\cite{grempel1982localization} as a model stemming from the study of
quantum chaos. In physics it
became quite popular as an exactly solvable example of the family of
incommensurate models, e.g. \cite{berry,new}, with spectral
transitions governed by arithmetics.  It was dubbed the
Maryland model
by B. Simon \cite{simm,CyconFroeseKirschSimon_book_1987} who  cited it as a textbook
example of dealing with small divisors. Already in  the early 80s the spectral theory of the Maryland model has been described for almost all values of $\alpha,\theta$ with arithmetic conditions on $\alpha$ \cite{simm,pf}. The region of the transition has been however difficult to handle. Recently, the Maryland model has become the first one for which the spectral decomposition has been determined for {\it all} parameters.
Define an index $\delta
(\alpha, \theta)\in [-\infty,\infty]$,
 \begin{equation}\label{Def.delta_1}
    \delta (\alpha, \theta)=\limsup_{n\rightarrow\infty} \frac{ \ln||q_n(\theta-1/2)||_{\mathbb{R}/\mathbb{Z}}+\ln q_{n+1}}{q_n},
 \end{equation}
 where $||x||_{\mathbb{R}/\mathbb{Z}}=\min_{\ell \in
   \mathbb{Z}}|x-\ell| $. Let  $\gamma_\lambda(e)$ be the Lyapunov
 exponent. remarkably, it is an explicit function, namely the solution of 
\begin{equation}\label{LEBiaoda}
     4\cosh \gamma_\lambda(e)=  \sqrt{ (2+e)^2 +\lambda^2} +\sqrt{ (2-e)^2 +\lambda^2},
 \end{equation}
%of the Maryland cocycle
Let
     %$\sigma_{p}(h)$,
$\sigma_{pp}(h)$,  $\sigma_{sc}(h)$ and
     $\sigma_{ac}(h)$ be the
% point spectrum (collection of all eigenvalues),
 pure point spectrum,
   singular   continuous spectrum, and absolutely continuous spectrum  of $h,$ respectively. Assume $\alpha\in \mathbb{R}\backslash \mathbb{Q}$ (otherwise
the spectrum is purely absolutely continuous),
$\lambda > 0$ (otherwise use
$h_{\lambda,\alpha,\theta}=h_{-\lambda,-\alpha,-\theta}$) and $
\theta\notin  1/2 + \alpha\mathbb{Z}+\mathbb{Z}$ (so the potential is
well defined).

\begin{theorem}\cite{jliu}\label{maryland}
For all $\alpha,\lambda,\theta$ as above , we have
\begin{enumerate}
  \item $\sigma_{sc}(h_{\lambda,\alpha,\theta})=\overline{\{e:\gamma_{\lambda}(e) <\delta (\alpha, \theta) \}}$
\item $\sigma_{pp}(h_{\lambda,\alpha,\theta})=\{e:\gamma_{\lambda}(e)
  \geq \delta (\alpha, \theta) \}$
\item $\sigma_{ac}(h_{\lambda,\alpha,\theta})=\emptyset.$
%Combing with the known fact  $\sigma_{ac}(h_{\lambda,\alpha,\theta})=\emptyset$,  we obtain the complete description
%of the three spectral types  of  Maryland Model.

%k_\lambda(e)\in \theta-1/2+   \alpha\mathbb{Z} +\mathbb{Z}\}$

%\item $\sigma_{p}(h_{\lambda,\alpha,\theta})$ and the set  $\{e:\gamma_{\lambda}(e) >\delta (\alpha, \theta) \}\cap Q(\theta)$    at
%most differ by  a (no more than) two-point set
%$\{e:\gamma_{\lambda}(e)=\delta (\alpha, \theta) \}$.

\end{enumerate}
\end{theorem} 
The proof is achieved through a new Gordon-type method, that allows to handle singular potentials in a sharp way, as well as a new technique for handling the cohomological equation in the regime of very small denominators.

\bibliographystyle{amsplain}

\end{document}